\documentclass{SCIS2024}

\makeatletter
\renewcommand{\maketag@@@}[1]{\hbox{\m@th\normalsize\normalfont#1}}%
\makeatother

\begin{document}

\ArticleType{RESEARCH PAPER}

\Year{2024}
\Month{}
\Vol{}
\No{}
\DOI{}
\ArtNo{}
\ReceiveDate{}
\ReviseDate{}
\AcceptDate{}
\OnlineDate{}

\title{BUPTCMCC-6G-CMG+: A GBSM-Based ISAC Standard Channel Model Generator}{ISAC channel simulator GBSM}

\author[1]{Changsheng ZHAO}{}
\author[1]{Jianhua ZHANG}{{jhzhang@bupt.edu.cn}}
\author[1]{Yuxiang ZHANG}{{zhangyx@bupt.edu.cn}}
\author[1]{Lei TIAN}{}
\author[1]{Heng WANG}{}
\author[1]{\\Hanyuan JIANG}{} 
\author[1]{Yameng LIU}{}
\author[1]{Wenjun CHEN}{}
\author[2]{Tao JIANG}{}
\author[2]{Guangyi LIU}{}

\AuthorMark{Author A}

\AuthorCitation{Author A, Author B, Author C, et al}

\address[1]{State Key Laboratory of Networking and Switching Technology, Beijing University of Posts and Telecommunications, \\Beijing {\rm 100876}, China}
\address[2]{Future Research Laboratory, China Mobile Research Institute,  Beijing {\rm 10053}, China}

\abstract{Integrated sensing and communication (ISAC) has been recognized as the key technology in the vision of the sixth generation (6G) era. With the emergence of new concepts in mobile communications, the channel model is the prerequisite for system design and performance evaluation. Currently, the 3rd Generation Partnership Project (3GPP) Release 19 is advancing the standardization of ISAC channel models. Nevertheless, a unified modeling framework has yet to be established. This paper provides a simulation diagram of ISAC channel modeling extended based on the Geometry-Based Stochastic Model (GBSM), compatible with existing 5G channel models and the latest progress in the 3GPP Rel-19 standardization. We first introduce the progress of the ISAC channel model standardization in general. Then, a concatenated channel modeling method is presented considering the team's standardization proposals, which is implemented on the BUPTCMCC-6G-CMG+ channel model generator. We validated the model in cumulative probability density function in statistical extension of angle and delay, and radar cross section (RCS). Simulation results show that the proposed model can realistically characterize the feature of channel concatenation and RCS within the ISAC channel.}

\keywords{Channel model generator, channel concatenation, ISAC channel, RCS, 3GPP standardization.}

\maketitle

\section{Introduction}
The sixth generation (6G) mobile communication technology is anticipated to enhance 5G communication capabilities roundly, including high capacity, low latency, and high reliability, while achieving native intelligence, integrated sensing and communication (ISAC), and ubiquitous connectivity \cite{1}. The boundary between the physical and digital realms is expected to become increasingly indistinct, offering users a ubiquitous and immersive experience \cite{2}. Concerns of communication systems will need to expand from information transmission to information acquisition and information processing\cite{3}. ISAC is the essential technology necessary to achieve the information paradigm transformation described above. ISAC technology can facilitate the implementation of diverse use cases, particularly in localization, tracking, imaging, and the detection of target type, velocity, and direction. These capabilities will enable applications in autonomous driving, industrial automation, digital twin reconstruction of physical environments, gesture detection, and activity recognition\cite{4,5,6,7,8}. The sensing information obtained from ISAC can facilitate the development of smart homes, factories, smart cities, and intelligent transportation systems. Meanwhile, ISAC technology introduces new performance metrics relative to traditional communication technologies, including detection probability, sensing capacity density, sensing accuracy, and resolution for velocity, angle, and delay \cite{9}.

The transmission medium between the transmitter and receiver establishes the performance limits of communication systems. The wireless channel is crucial for the research and evaluation of wireless communication technologies \cite{10,11,12}. The performance evaluation of the ISAC system introduces new requirements for channel models, particularly regarding sensing modes and scenarios, sensing use cases, and channel modeling methodologies. Considering sensing modes, current channel models usually utilize Bi-static modes for communication. Sensing applications necessitate considering the case of Mono-static channel models for self-sensing. Thus, ISAC channel models should incorporate both Mono-static and Bi-static sensing modes. Secondly, concerning sensing use cases involving sensing targets, ISAC applications expand the evaluation criteria to include target detection, localization, identification, and tracking. Examples encompass intrusion detection, unmanned aerial vehicle (UAV) tracking, vehicle-to-everything (V2X) communications, and the identification of diverse target types and sizes. That requires utilizing the scattering properties of radar cross section (RCS), polarization characteristics, delay, and Doppler information of electromagnetic waves as they interact with the target. Consequently, channel models should integrate the RCS characteristics of various targets. Moreover, for modeling methodologies, it is imperative to account for the concatenated characteristics of channels interacting with the target while coupling with existing standard communication channel models \cite{13,14,15,16}.

So far, the standardization of ISAC channel models by the 3rd Generation Partnership Project (3GPP) has held five conferences from RAN1 \#116 to RAN1 \#118-bis, wherein ISAC scenarios, types of sensing targets, sensing modes, and general channel model frameworks have been defined. Recent studies have demonstrated the latest progress for 3GPP ISAC channel modeling standardization \cite{17}. The defined scenarios encompass traditional urban micro (UMi), urban macro (UMa), rural macro (RMa), and indoor hotspot (InH) scenarios, in addition to V2X, UAV, and industrial Internet of things (IIoT) scenarios. Accordingly, sensing targets are classified into five categories: humans, vehicles, UAVs, automated guided vehicles (AGVs), and obstacles, as shown in Table \ref{tab1}. Sensing modes are categorized into Bi-static and Mono-static, which are further subdivided into six specific use cases based on whether the sensing device is a user terminal (UT) or a base station (BS): BS-UT-Bi, BS-BS-Bi, UT-UT-Bi, UT-BS-Bi, UT-UT-Mono, and BS-BS-Mono \cite{18}. In the framework of channel modeling, the ISAC channel is defined as the combination of the target channel and the background channel. The target channel includes all channel components that interact with targets, whereas the background channel comprises all other elements that do not engage with targets.

For ISAC channel simulation, a channel model that separately generates communication and sensing clusters based on the 3GPP model and utilizes sensing echoes for sensing has been proposed \cite{19}. Some studies introduce the concept of shared clusters to explain the phenomenon of scattering body reuse between the communication and sensing channels \cite{5, 20, 21, 22}. Some studies focus on deterministic ISAC channel modeling \cite{23, 24}. However, these studies do not align with the target and background channel modeling framework or the concatenated channel method in 3GPP Rel-19. Most of these approaches use Mono-static echoes for sensing and do not support Bi-static sensing modes.
\begin{table}[!t] 
\footnotesize
\caption{3GPP ISAC Standardization Focused Scenarios }
\label{tab1}
\tabcolsep 40pt 
\begin{tabular*}{\textwidth}{cc}
\toprule
  Sensing targets & Scenarios \\\hline
  UAVs & RMa-AV, UMa-AV, UMi-AV \\
  Humans indoors & InF, InH, UMi, UMa, RMa \\
  Automotive vehicles & Highway, Urban grid, UMi, UMa, RMa\\
  Automated guided vehicles & InF \\
  Objects creating hazards on road/railways & Highway, Urban grid, HST\\
\bottomrule
\end{tabular*}
\end{table}

To the authors' knowledge, no unified ISAC channel simulation framework compatible with the latest 3GPP standardization progress has been proposed. Considering that standardization is primarily driven by industry and the existing standardization framework has not yet been fully finalized, proposing a unified channel expression that integrates key aspects discussed in the RAN1 agenda such as RCS modeling, the concatenated channel methodology, and micro-Doppler and macro-Doppler characteristics prior to 3GPP's finalization is crucial for understanding the progress in ISAC standardization and the practical implementation of ISAC channel models. Given this, we present our solution, which is grounded in our team's contributions to ISAC standardization and our understanding of standardized channel simulation. The contributions and innovations of this paper are outlined as follows:

\begin{itemize}
\item[(a)]
Different from existing ISAC channel simulation models \cite{19,20,21,22,23,24}, we propose a unified ISAC channel simulation framework based on the 3GPP Rel-19 ISAC channel fundamental framework, where the ISAC channel is the combination of the target and background channels. This framework maximizes the use and compatibility with existing 5G standardized channel models. This work integrates our team's contributions to the standardization process, presenting a solution that aligns with the latest standardized progress and consensus in ISAC channel modeling \cite{17}.
\item[(b)]
The effective ISAC channel simulation method has been validated.
\begin{itemize}
    \item  We compared and analyzed the down-selection simulation results for the 8 concatenation options presented in the RAN1 \#118-bis meeting, explaining the mechanisms behind the results. Based on it, we propose a simulation method for ISAC concatenated channels.
    \item  A unified RCS simulation model is introduced, which considers the progress of the RAN1 \#116-118bis meetings and incorporates our team's RCS measurement results. This model applies to human, vehicle, and UAV targets, accounting for factors such as large-scale, small-scale, statistical, and deterministic components.
    \item  We propose a solution for combining the target and background channels, compatible with two existing standardized approaches: one where the target channel is included in the background channel and the other where the target channel is directly added to the background channel.
    \item Key performance evaluation metrics for ISAC are provided, along with a brief analysis of the underlying mechanisms for each metric.
\end{itemize}

\end{itemize}

The paper is organized as follows. Section $2$ describes the simulation architecture in detail. Section $3$ reports on evaluation metrics for ISAC. Then, Section $4$ presents the simulation results for both ISAC channel concatenation and RCS. Section $5$ concludes the paper.
\section{Simulation Architecture}
\begin{figure}[!h]
\centering
\includegraphics[width=6.5in]{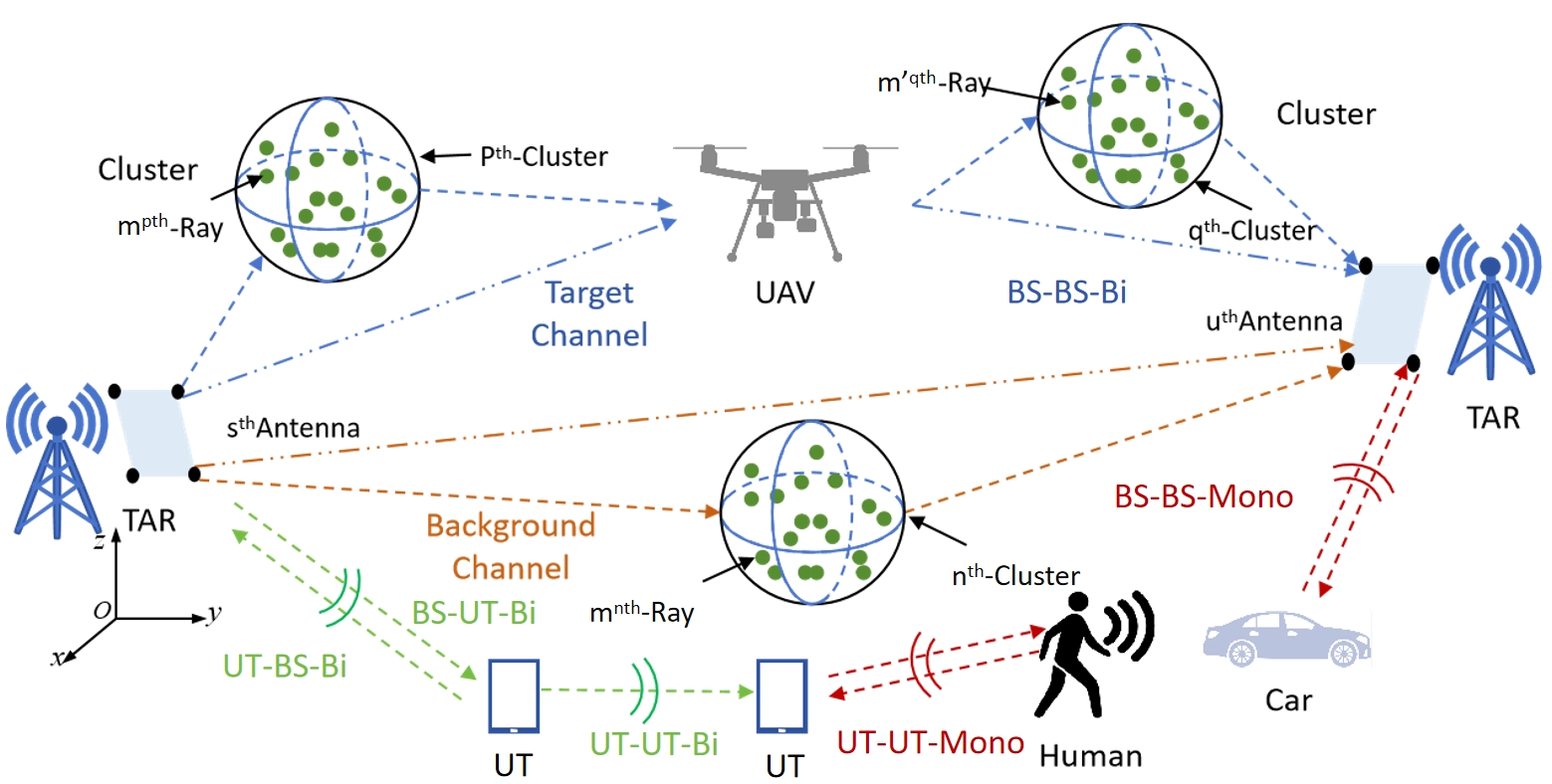}
\caption{A geometric schematic of ISAC channel paradigm}
\label{fig1}
\end{figure}

 Building upon the current advancements in 3GPP standardization efforts and the GBSM simulation framework, this paper proposes an ISAC channel simulation framework designed for future standardization, which is integrated into the BUPTCMCC-6G-CMG+ channel model generator. The key characteristics of ISAC channels—including concatenated features, RCS polarization, RCS scattering properties, and the combination of target and background channels, are considered to establish a unified model for ISAC channel simulation. In this model, the ISAC channel is divided into two components: the target channel and the background channel. The geometric schematic of the model is shown in Figure \ref{fig1}.

The target channel is further classified into Tx-target and target-Rx sub-channels. For Bi-static, the target channel is constructed by cascading the Tx-target and target-Rx sub-channels, with parameters for each sub-channel generated according to the existing 3GPP models. For Mono-static, the Tx-target link is generated using the 3GPP model, whereas the target-Rx components are derived by utilizing channel reciprocity.
Furthermore, the background channel is generated directly using the 3GPP statistical channel model for the Bi-static sensing mode. As to the Mono-static sensing mode, the background channel can be derived from the 3GPP method, which has limited related work, or extracted through additional measurements to obtain the necessary channel model parameters.

\begin{figure}[!h]
\centering
\includegraphics[width=6in]{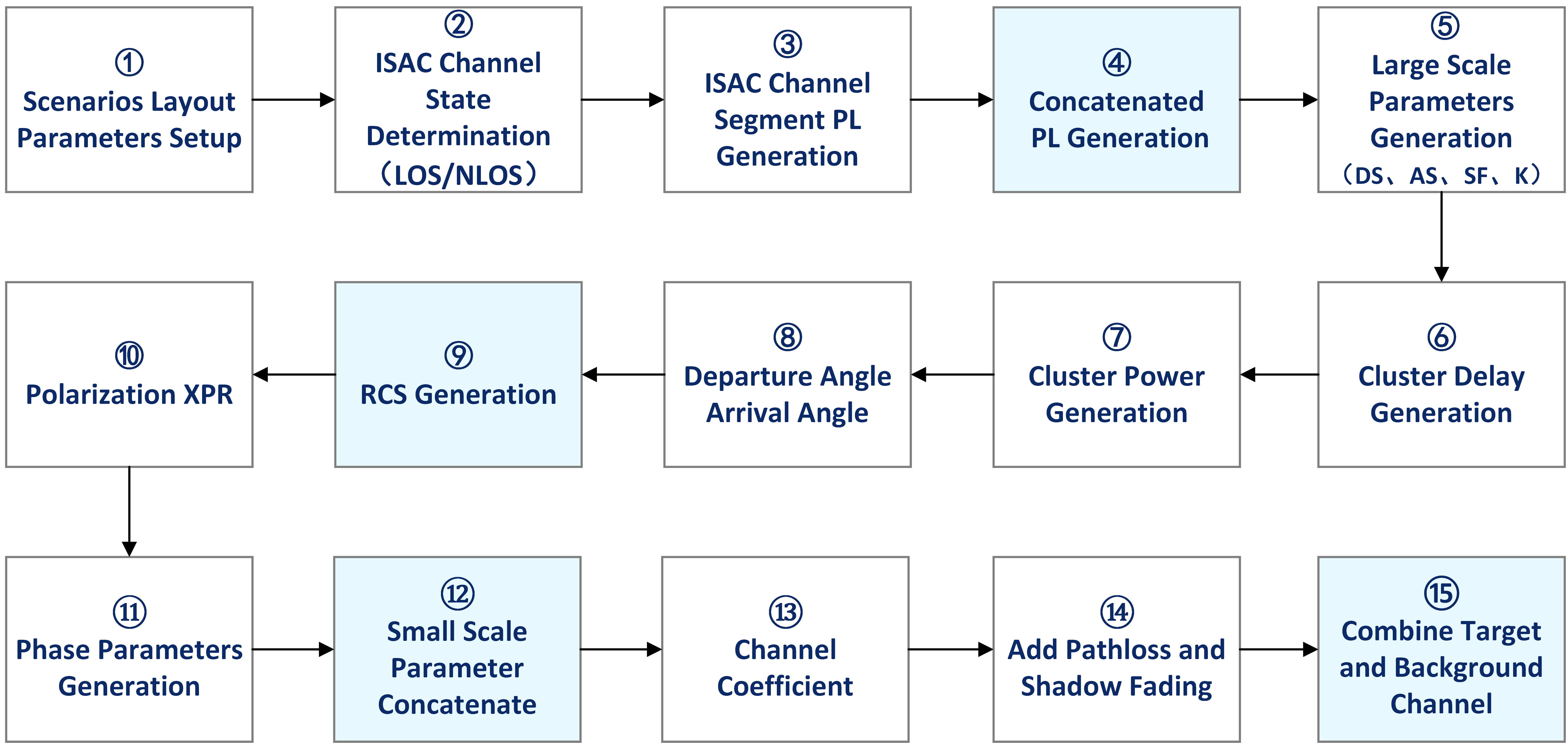}
\caption{The simulation steps for generating ISAC channel coefficients. (changes compared to the GBSM model are highlighted)}
\label{fig2}
\end{figure}
Figure \ref{fig2} depicts the proposed simulation process extended by the 3GPP channel model\cite{13,14,15}. In Step 1, users can perform general configurations, including antenna configuration, scenario settings, target types, and BS, UT, and target positions. Steps 2 to 5 are dedicated to large-scale parameter generation, while Steps 6 to 12 concentrate on small-scale parameter generation. Finally, Steps 13 and 14 are responsible for generating the channel coefficients. Steps 4, 9, and 12 correspond to large-scale parameter concatenation, RCS generation, and small-scale parameter concatenation, respectively. Step 15 refers to the combination of target and background channels.

\subsection{Concatenated channel modeling}
 In this section, we propose a parameter-concatenated method for ISAC channel simulation with scenario layouts, path loss formulas, line-of-sight (LOS) probabilities, and large-scale parameters defined in the 3GPP channel model standards TR38.901, TR36.777, and TR37.885. Specifically, TR38.901 is used for scenarios involving human and AGV sensing targets, TR36.777 is designated for UAV scenarios, and TR37.885 is applied to V2X scenarios. The concatenated simulation method comprises two main components: large-scale parameters concatenation and small-scale parameters concatenation.
\subsubsection{Large scale parameters concatenation}
 As illustrated in Figure \ref{fig2}, the simulation starts with the establishment of various parameters, including the simulation scenario, target type, target position, velocity and direction, the number and location of base stations and terminals, antenna patterns, orientation of both base stations and terminals, system center frequency, speed and direction of the terminals. For the configured scenario, the LOS probability and path loss for the Tx-target and target-Rx links are computed individually. The corresponding mean value of RCS is generated for the target. The concatenation path loss can be calculated as
\begin{equation}
\label{eq_PL}
PL_{target}(d_1,d_2,f)=PL_{1}(d_1,f)+PL_{2}(d_2,f)+10\log{\frac{c^2}{4 \pi f^2}} - 10\log{\overline{\sigma}_{RCS}},
\end{equation}
where $d_1$ and $d_2$ represent the distances of the Tx-target and target-Rx, respectively. $PL_{1}(\cdot)$ and $PL_{2}(\cdot)$ represent the path loss of the Tx-target and target-Rx links, respectively. $f$ denotes the center frequency of the simulation. $c$ is the speed of light with $3 \times 10^8$ m/s. $\overline{\sigma}_{RCS}$ is the mean RCS.
  
\subsubsection{Small scale parameters concatenation}
In this section, we derive and present a unified function for the ISAC target channel. This function is compatible with the concatenated channel methodology outlined in the 3GPP RAN1 ISAC channel modeling issue, incorporating key parameters such as RCS small-scale variations, target polarization, and Doppler. The unified function for each receiver and transmitter antenna element pair u and s in LOS+LOS condition is given in (\ref{eq_overall}), while the equations for the remaining LOS+NLOS, NLOS+LOS, and NLOS+NLOS conditions can be obtained by simplifying \ref{eq_overall}.
\begin{align}
\label{eq_overall}
H_{u,s}^{\mathrm{target}}(\tau,t)&=\sqrt{\frac{K_P}{K_P+1}}\sqrt{\frac{K_Q}{K_Q+1}}H_{u,s}^{\mathrm{LOS,LOS}}(t)\delta\left(\tau-\tau_{1,1}\right)+\sqrt{\frac{K_P}{K_P+1}}\sqrt{\frac{1}{K_Q+1}}H_{u,s}^{\mathrm{LOS,NLOS}}(\tau,t)\nonumber\\ &+\sqrt{\frac{1}{K_P+1}}\sqrt{\frac{K_Q}{K_Q+1}}H_{u,s}^{\mathrm{NLOS,LOS}}(\tau,t)+\sqrt{\frac{1}{K_P+1}}\sqrt{\frac{1}{K_Q+1}}H_{u,s}^{\mathrm{NLOS,NLOS}}(\tau,t).
\end{align}
Here, $K_P$ and $K_Q$ represent the Ricean K-factors for the Tx-target and target-Rx links, respectively, which are generated in Step 5 and converted to a linear scale. $\delta(\cdot)$ denotes the Dirac delta function. $\tau_{1,1}$ is the sum of the delays of LOS rays in Tx-target and target-Rx links. The terms $H_{u,s}^{LOS,LOS}$, $H_{u,s}^{LOS,NLOS}$, $H_{u,s}^{NLOS,LOS}$, $H_{u,s}^{NLOS,NLOS}$, and correspond to the matching components for LOS-LOS path, LOS-NLOS multipath, NLOS-LOS multipath, and NLOS-NLOS multipath of the target channel coefficients respectively. Their detailed expressions are provided in (\ref{eq3}), (\ref{eq4}), (\ref{eq5}), and (\ref{eq6}).

\begin{align}
\label{eq3}
    H_{u,s}^{LOS,LOS}(t)=&\begin{bmatrix}F_{rx,u,\theta}(\theta_{Q,LOS,ZOA},\phi_{Q,LOS,AOA})\\F_{rx,u,\phi}(\theta_{Q,LOS,ZOA},\phi_{Q,LOS,AOA})\end{bmatrix}^T\textbf{P}_{LOS,LOS}\begin{bmatrix}F_{tx,s,\theta}(\theta_{P,LOS,ZOD},\phi_{P,LOS,AOD})\\F_{tx,s,\phi}(\theta_{P,LOS,ZOD},\phi_{P,LOS,AOD})\end{bmatrix}\nonumber\\ &exp(j\frac{2\pi}{\lambda}\hat{r}_{rx,LOS}^T\cdot\overline{d}_{rx,u})exp(j\frac{2\pi}{\lambda}\hat{r}_{tx,LOS}^T\cdot\overline{d}_{tx,s})exp(j2\pi f_{d,1,1}t)\sqrt{\sigma_{1,1}}.
\end{align}

\begin{align}
\label{eq4}
    H_{u,s}^{LOS,NLOS}(\tau,t)=&\sqrt{\frac{1}{\sum_{q=1}^{Q}P_{q}}}\sum_{q}^{Q}\sum_{m^{\prime}}^{M^{\prime}}\sqrt{\frac{P_{q}}{M^{\prime}}}\begin{bmatrix}F_{rx,u,\theta}(\theta_{q,m^{\prime},ZOA},\phi_{q,m^{\prime},AOA})\\F_{rx,u,\phi}(\theta_{q,m^{\prime},ZOA},\phi_{q,m^{\prime},AOA})\end{bmatrix}^{T} \nonumber \\&\mathbf{P}_{LOS,NLOS}\begin{bmatrix}F_{tx,s,\theta}(\theta_{P,LOS,ZOD},\phi_{P,LOS,AOD})\\F_{tx,s,\phi}(\theta_{P,LOS,ZOD},\phi_{P,LOS,AOD})\end{bmatrix}exp(j\frac{2\pi}{\lambda}\hat{r}_{rx,q,m^{\prime}}^T\cdot\bar{d}_{rx,u}) \nonumber \\&exp(j\frac{2\pi}{\lambda}\hat{r}_{tx,LOS}^T\cdot\bar{d}_{tx,s})exp(j2\pi f_{d,1,q,m^{\prime}}t)\delta(\tau-\tau_{P,1}-\tau_{q,m^{\prime}})\sqrt{\sigma_{1,q,m^{\prime}}}.
\end{align}

\begin{align}
\label{eq5}
    H_{u,s}^{NLOS,LOS}(\tau,t)=&\sqrt{\frac{1}{\sum_{p=1}^{P}P_{p}}}\sum_{p}^{P}\sum_{m}^{M}\sqrt{\frac{P_{p}}{M}}\begin{bmatrix}F_{rx,u,\theta}(\theta_{Q,LOS,ZOA},\phi_{Q,LOS,AOA})\\F_{rx,u,\phi}(\theta_{Q,LOS,ZOA},\phi_{Q,LOS,AOA})\end{bmatrix}^{T}\nonumber\\&\mathbf{P}_{NLOS,LOS}\begin{bmatrix}F_{tx,s,\theta}(\theta_{p,m,ZOD},\phi_{p,m,AOD})\\F_{tx,s,\phi}(\theta_{p,m,ZOD},\phi_{p,m,AOD})\end{bmatrix}exp(j\frac{2\pi}{\lambda}\hat{r}_{rx,LOS}^{T}\cdot\bar{d}_{rx,u})\nonumber\\&exp(j\frac{2\pi}{\lambda}\hat{r}_{tx,p,m}^{T}\cdot\bar{d}_{tx,s})exp(j2\pi f_{d,p,m,1}t)\delta(\tau-\tau_{p,m}-\tau_{Q,1})\sqrt{\sigma_{p,m,1}}.
\end{align}

\begin{align}
\label{eq6}
\begin{aligned}H_{u,s}^{NLOS,NLOS}(\tau,t)=&\sqrt{\frac{1}{\sum_{p=1}^{P}P_{p}}}\sqrt{\frac{1}{\sum_{q=1}^{Q}P_{q}}}\sum_{p}^{P}\sum_{m}^{M}\sum_{q}^{Q}\sum_{m^{\prime}}^{M^{\prime}}\sqrt{\frac{P_{p}}{M}}\sqrt{\frac{P_{q}}{M^{\prime}}}\begin{bmatrix}F_{rx,u,\theta}(\theta_{q,m^{\prime},ZOA},\phi_{q,m^{\prime},AOA})\\F_{rx,u,\phi}(\theta_{q,m^{\prime},ZOA},\phi_{q,m^{\prime},AOA})\end{bmatrix}^{T}\\&\mathbf{P}_{NLOS,NLOS}\begin{bmatrix}F_{tx,s,\theta}(\theta_{p,m,ZOD},\phi_{p,m,AOD})\\F_{tx,s,\phi}(\theta_{p,m,ZOD},\phi_{p,m,AOD})\end{bmatrix}exp(j\frac{2\pi}{\lambda}\hat{r}_{rx,q,m^{\prime}}^T\cdot\bar{d}_{rx,u})\\&exp(j\frac{2\pi}{\lambda}\hat{r}_{tx,p,m}^T\cdot\bar{d}_{tx,s})exp(j2\pi f_{d,p,m,q,m^{\prime}}t)\delta(\tau-\tau_{p,m}-\tau_{q,m^{\prime}})\sqrt{\sigma_{p,m,q,m^{\prime}}}.\end{aligned}
\end{align}

Here, $P$ and $Q$ are the number of clusters in the Tx-target and target-Rx links, respectively. $\tau_{P,1}$ and $\tau_{Q,1}$ represent the LOS ray's delay of Tx-target and target-Rx links, respectively. $\tau_{p,m}$ and $\tau_{q,m'}$ represent the delay of the $m$-th ray within the $p$-th cluster in the Tx-target link and the delay of the $m'$-th ray within the $q$-th cluster in target-Rx link, respectively. Note that the delays referenced in this paper align with the relative delays specified in 3GPP TR 38.901 Section 7.5. For practical ISAC channel simulations, absolute delays need to be computed to evaluate sensing performance. For LOS condition, absolute delays can be calculated geometrically by $\frac{d_{3D}}{c}$, where $d_{3D}$ is the 3D distance. For NLOS condition, Release 19 discussions are ongoing; meanwhile, Section 7.6.9 of TR 38.901 may serve as a provisional baseline. The actual propagation delay for ISAC channel can then be obtained by superimposing the absolute delay onto the relative delays generated in this work. $\lambda$ is the wavelength corresponding to the center carrier frequency. $M$ and $M'$ are the number of rays in each cluster in the Tx-target and target-Rx links, respectively. $F_{rx,u,\theta}(\cdot)$ and $F_{rx,u,\phi}(\cdot)$ represent the field components of the receiving antenna $u$ in the zenith and azimuth directions, respectively. $\theta$ and $\phi$ represent the zenith and the azimuth angle, respectively.  $F_{tx,s,\theta}(\cdot)$ and $F_{tx,s,\phi}(\cdot)$ represent the field components of the transmitting antenna $s$ in the zenith and azimuth directions, respectively. $\sigma$ represents the RCS value related to angular terms, the detailed definition of which can be found in Section 2.2.2. $P_p$ and $P_q$ are the power of the $p$-th cluster in the Tx-target link and the power of the $q$-th cluster in the target-Rx link, respectively. $\overline{d}_{rx,u}$ and $\overline{d}_{tx,s}$ represent the position vectors of the receiving and transmitting antenna elements, respectively. $\hat{r}_{rx,q,m'}$ and $\hat{r}_{tx,p,m}$ are the unit vectors in spherical coordinate, with the calculation expressions as follows:
\begin{equation}
\label{eq7}
\left.\hat{r}_{rx,q,m^{\prime}}=\left[\begin{array}{c}\sin\theta_{q,m^{\prime},ZOA}\cos\theta_{q,m^{\prime},AOA}\\\sin\theta_{q,m^{\prime},ZOA}\sin\theta_{q,m^{\prime},AOA}\\\cos\theta_{q,m^{\prime},ZOA}\end{array}\right.\right],\hat{r}_{tx,p,m}=\begin{bmatrix}\sin\theta_{p,m,ZOD}\cos\theta_{p,m,AOD}\\\sin\theta_{p,m,ZOD}\sin\theta_{p,m,AOD}\\\cos\theta_{p,m,ZOD}\end{bmatrix}.
\end{equation}

 $f_d$ is the Doppler frequency for a target including both macro-Doppler and micro-Doppler, which can be modeled using a unified formula as:
 
\begin{equation}
\label{eq8}
f_d=\frac{\hat{r}_{rx,q,m'}^T\cdot\overline{\nu}_{rx}+\hat{r}_{sp,q,m'}^T\cdot\overline{\nu}_{sp}+\hat{r}_{tx,p,m}^T\cdot\overline{\nu}_{tx}+\hat{r}_{sp,p,m}^T\cdot\overline{\nu}_{sp}}{\lambda},
\end{equation}
where, $\hat{r}_{rx,q,m'}^{T}$ is the spherical unit vector at the receiver for the link from Rx to the scattering point. $\hat{r}_{sp,q,m'}^{T}$ is the spherical unit vector at the scattering point for the link from the scattering point to Rx. $\hat{r}_{tx,p,m}^{T}$ is the spherical unit vector at the receiver for the link from Tx to the scattering point. $\hat{r}_{sp,p,m}^{T}$ is the spherical unit vector at the scattering point for the link from the scattering point to Tx. $\overline{v}_{tx}$, $\overline{v}_{rx}$, and $\overline{v}_{sp}$ are the velocity of the Tx, Rx, and scattering point respectively.   $\overline{v}_{sp}$ can include macro-Doppler and micro-Doppler motion, $\overline{v}_{sp}=\overline{v}_{macro}+\overline{v}_{micro}$.

$\mathbf{P}_{LOS,LOS}$, $\mathbf{P}_{LOS,NLOS}$, $\mathbf{P}_{NLOS,LOS}$, and $\mathbf{P}_{NLOS,NLOS}$ are polarization matrices correspond to the LOS-LOS, LOS-NLOS, NLOS-LOS, and NLOS-NLOS path of the target channel, respectively.

\begin{equation}
\label{eq9}
\mathbf{P}_{LOS,LOS}=\begin{bmatrix}exp(j\Phi_Q^{LOS})&0\\0&-exp(j\Phi_Q^{LOS})\end{bmatrix}\mathbf{S_1}\begin{bmatrix}exp(j\Phi_P^{LOS})&0\\0&-exp(j\Phi_P^{LOS})\end{bmatrix},
\end{equation}

\begin{equation}
\label{eq10}
\mathbf{P}_{LOS,NLOS}=\begin{bmatrix}exp(j\Phi_{q,m^{\prime}}^{\theta\theta})&\sqrt{\kappa_{q,m'}^{-1}}exp(j\Phi_{q,m^{\prime}}^{\theta\phi})\\\sqrt{\kappa_{q,m^{\prime}}^{-1}}exp(j\Phi_{q,m^{\prime}}^{\phi\theta})&exp(j\Phi_{q,m^{\prime}}^{\phi\phi})\end{bmatrix}\mathbf{S_2}\begin{bmatrix}exp(j\Phi_P^{LOS})&0\\0&-exp(j\Phi_P^{LOS})\end{bmatrix},
\end{equation}

\begin{equation}
\label{eq11}
\mathbf{P}_{NLOS,LOS}=\begin{bmatrix}exp(j\Phi_{Q}^{LOS})&0\\0&-exp(j\Phi_{Q}^{LOS})\end{bmatrix}\mathbf{S_3}\begin{bmatrix}exp(j\Phi_{p,m}^{\theta\theta})&\sqrt{\kappa_{p,m}^{-1}}exp(j\Phi_{p,m}^{\theta\phi})\\\sqrt{\kappa_{p,m}^{-1}}exp(j\Phi_{p,m}^{\phi\theta})&exp(j\Phi_{p,m}^{\phi\phi})\end{bmatrix},
\end{equation}

\begin{equation}
\label{eq12}
\mathbf{P}_{NLOS,NLOS}=\begin{bmatrix}exp(j\Phi_{q,m'}^{\theta\theta})&\sqrt{\kappa_{q,m'}^{-1}}exp(j\Phi_{q,m'}^{\theta\phi})\\\sqrt{\kappa_{q,m'}^{-1}}exp(j\Phi_{q,m'}^{\phi\theta})&exp(j\Phi_{q,m'}^{\phi\phi})\end{bmatrix}\mathbf{S_4}\begin{bmatrix}exp(j\Phi_{p,m}^{\theta\theta})&\sqrt{\kappa_{p,m}^{-1}}exp(j\Phi_{p,m}^{\theta\phi})\\\sqrt{\kappa_{p,m}^{-1}}exp(j\Phi_{p,m}^{\phi\theta})&exp(j\Phi_{p,m}^{\phi\phi})\end{bmatrix}.
\end{equation}
Where, $\kappa_{p,m}$ and $\kappa_{q,m'}$ are the cross-polarization power ratio (XPR) in the Tx-target and target-Rx link, respectively. $\Phi_{P}^{LOS}=exp(-\frac{2\pi}{\lambda}d_{3D}^{Tx-target})$ and $\Phi_{Q}^{LOS}=exp(-\frac{2\pi}{\lambda}d_{3D}^{target-Rx})$ represent the phase of the LOS ray of Tx-target and target-Rx links, respectively. Here, $d_{3D}^{Tx-target}$ and $d_{3D}^{target-Rx}$ are the three dimension (3D) distance of Tx-target and target-Rx links, respectively. $\mathbf{S_1}$, $\mathbf{S_2}$, $\mathbf{S_3}$, and $\mathbf{S_4}$ represent the polarization effects induced by the target for the LOS+LOS, LOS+NLOS, NLOS+LOS, and NLOS+NLOS link conditions, all of which can be calculated following (\ref{eq_S}).

During the 3GPP RAN1 \#116 meeting, RAN1 defined the ISAC channel as consisting of the target and background channels. However, the methodology for generating the combined ISAC channel is subject to further study. The relationship between the background channel component and the environment channel in the absence of sensing targets remains an unresolved issue. To tackle this issue, we are introducing a new parameter called the coupling factor\cite{25}, to integrate the path loss of both the target and background channels. Consequently, the combined path loss of the ISAC channel can be expressed as follows:
\begin{equation}
    \label{eq_norm}
    PL_{ISAC} = PL_{target} + O_{isac}\cdot PL_{backgraound}^{standardization},
\end{equation}
where the $ PL_{ISAC}$ and $PL_{target}$ represent the linear path loss values of the ISAC channel and target channel, respectively. The $PL_{background}^{standardization}$ represents the environment channel, i.e., the initial background channel with no sensing target, and it is important to note that $PL_{background}^{standardization}$  is not the same as the defined $PL_{backgraound}$. The latter refers to the path loss of the channel components after the target channel components have been removed. Given the similarities and compatibility with the communication channel, it is advisable to reuse the path loss model from 3GPP TR 38.901. The values of $O_{isac}$ need further study with more measurement results.

\subsection{RCS modeling}
Radar systems leverage the scattered power of targets for sensing, and the accurate modeling of the RCS of these targets directly influences the performance evaluation of sensing capabilities. First, it is essential to introduce the concept of the three-dimensional resolution cell. A resolution cell is defined as the volume that is illuminated and scattered at a specific moment in time, characterized by two dimensions: angle and distance. It can be expressed as follows
\begin{equation}
\label{eq_resolution}
V = \frac{\Omega R^2 c \tau}{2},
\end{equation}
where $\frac{c\tau}{2}$ represents the distance resolution cell, and $\tau$ is the pulse width. $\Omega$ is the antenna beam solid angle, and $R$ is the distance from the radar receiver to the target. If the target's size is contained within the volume $V$, the target is considered a point target.
For the five types of sensing targets defined by 3GPP UAVs, humans, long-distance vehicles, AGVs, and small road obstacles modeling as point targets is appropriate. However, for targets that exceed the resolution cell and possess irregular shapes, such as close-range vehicles and AGVs, large obstacles, and humans in gesture recognition scenarios, multi-point RCS modeling may be considered.

\subsubsection{RCS polarization}
The scattering characteristics of a target are influenced by the polarization of the incident field. When the antenna is linearly polarized, the incident linearly polarized wave on the target surface can be decomposed into vertical and horizontal polarization components. The polarization components of the scattered field from the target, as received by the receiving antenna, can be expressed as
\begin{equation}
\label{eq_RCSpolar}
\begin{bmatrix}E_V^R\\E_H^R\end{bmatrix}=\mathbf{S}\begin{bmatrix}E_V^T\\E_H^T\end{bmatrix}.
\end{equation}
Here, $E_{V}^{T}$ and $E_{H}^{T}$ represent the vertical and horizontal polarization electric field component at the transmitter, respectively. $E_{V}^{R}$ and $E_{H}^{R}$ represent the vertical and horizontal polarization electric field component at the receiver, respectively. It should be noted that the electric field in (\ref{eq_RCSpolar}) only considers the effect of polarization, and the modeling of RCS in this paper is independent of polarization. The target polarization matrix $\textbf{S}$ is defined as follows:
\begin{equation}
\label{eq_S}
\mathbf{S}=\begin{bmatrix}\alpha_{V,V}exp\left(j\Phi_{sp}^{\theta\theta}\right)&\alpha_{V,H}exp\left(j\Phi_{sp}^{\theta\phi}\right)\\\alpha_{H,V}exp\left(j\Phi_{sp}^{\phi\theta}\right)&\alpha_{H,H}exp\left(j\Phi_{sp}^{\phi\phi}\right)\end{bmatrix},
\end{equation}
where, the initial random phase $\{\phi_{sp}^{\theta\theta},\phi_{sp}^{\theta\phi},\phi_{sp}^{\phi\theta},\phi_{sp}^{\phi\phi}\}$ is uniformly distributed within $(-\pi,\pi)$. For $\{\alpha_{V,V},\alpha_{V,H},\alpha_{H,V},\alpha_{H,H}\}$, there are three primary perspectives within RAN1: $\{1,0,0,1\}$, $\{\alpha_1,\alpha_2,\alpha_3,\alpha_4\}$, and $\{1,\alpha_2,\alpha_3,1\}$.  Here, $\alpha_1$, $\alpha_2$, $\alpha_3$, and $\alpha_4$ are variables generated for each path, and their specific values remain subject to further study. 

\subsubsection{RCS modeling based Measurement}
Drawing upon the comprehensive analysis of measurements that depend on frequency, distance, and target type conducted by our team in \cite{26,27}, the following conclusions regarding RCS modeling can be established:
\begin{itemize}
    \item The RCS of the UAV is independent of angle and demonstrates characteristics akin to omnidirectional scattering. Furthermore, the measurement results obtained from both Mono-static and Bi-static modes closely approximate each other.
    \item The RCS model for the UAV is represented as a normal distribution in both Mono-static and Bi-static sensing modes. The mean defines the deterministic component, while the variance characterizes the stochastic component.
    \item The RCS of a human is angle-independent and displays characteristics akin to omnidirectional scattering. Furthermore, the measurement results obtained from Mono-static and Bi-static modes closely align with each other.
    \item The RCS model for humans is represented as a normal distribution in both Mono-static and Bi-static sensing modes. The mean value characterizes the deterministic component, while the variance accounts for the stochastic component.
\end{itemize}
For different types of sensing targets, our team has proposed a unified RCS modeling framework in RAN1 \#118bis meeting \cite{28}:
\begin{equation}
    \label{eq_RCS}
    RCS = A \times B_1 \times B_2,
\end{equation}
where $A$ component is included in the large scale and is a deterministic value. The $B_1$ component is included in the small scale and is a deterministic angle-dependent value or scattering pattern based on measurements. The $B_2$ component is a log-normal distribution with a mean and variance. $A$ is used as $\overline{\sigma}_{RCS}$ in (\ref{eq_PL}), while $B_1$ and $B_2$ are used as $\sigma$ in (\ref{eq3}), (\ref{eq4}), (\ref{eq5}), (\ref{eq6}). Modeling the RCS of a UAV based on this framework, since its RCS is angle-dependent, the $A$ component is set to the mean or peak value, or simply set to 1, etc., the $B_1$ component is a function related to the angle, and  $B_2$ is a random value independent of the angle, representing the fluctuation of RCS. This model framework can effectively model the RCS of sensing targets with different scattering patterns. In section 4, we present the measured results of the human target RCS and the simulation fitting results based on the proposed framework, which are used to comparatively validate the feasibility of the model.

Different from the human body and small UAV, vehicle RCS and large UAV exhibit spiking characteristics that render them unsuitable for modeling with omnidirectional patterns.
The RCS of humans and small UAVs can be modeled using $A$ and $B_2$ with a log-normal distribution, in which the mean \( \mu \)  and variance \( \sigma \) are determined from empirical measurements. The RCS of vehicles and large UAVs displays four distinct peaks in its angular distribution, highlighting the need for further investigation into $B_1$ and simulation methodologies. 

\begin{figure}[!h]
\centering
\includegraphics[width=3in]{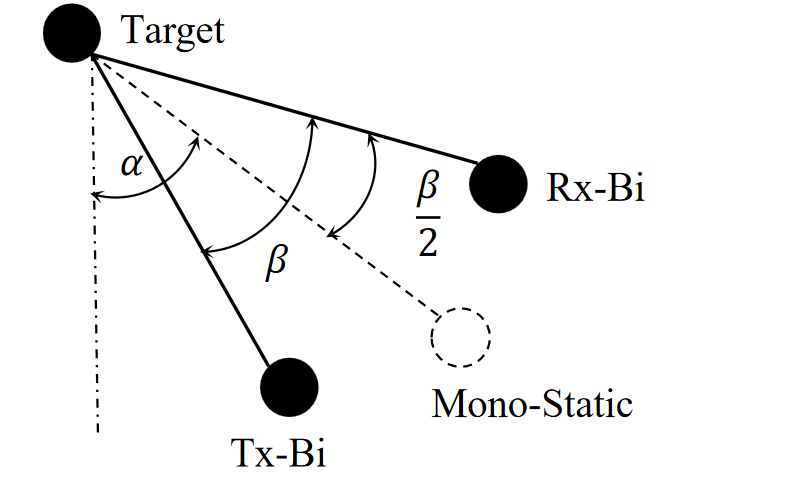}
\caption{Schematic diagram of mono-static sensing and bi-static sensing}
\label{fig3}
\end{figure}

The investigation of both Mono-static and Bi-static RCS is of equal importance, and their relationship warrants further discussion in RAN1. As illustrated in Figure \ref{fig3}, Mono-static sensing and Bi-static sensing are presented. Notably, Mono-static sensing can be regarded as a special case of Bi-static sensing when $\beta=0$. The equivalence relationship between Mono-static and Bi-static RCS was initially explored in radar research. Kell proposed a general scattering center-based Mono-static-to-Bi-static equivalence theorem (MBET) \cite{29}. This theorem demonstrates that the Bi-static RCS can be approximated by the Mono-static RCS measured at a frequency scaled by $\cos{\frac{\beta}{2}}$ on the bisector of the Bi-static angle $\beta$, as expressed by $\sigma\left(fcos\frac{\beta}{2},\alpha,\alpha\right)\approx\sigma\left(f,\alpha-\frac{\beta}{2},\alpha+\frac{\beta}{2}\right)$. where $\alpha$ is the receiver orientation angle, and $f$ is the carrier frequency. When the Bi-static angle $\beta$ is small, the Bi-static RCS can be approximated by the Mono-static RCS as follows:

\begin{equation}
\label{eq_RCSMono}
RCS_{Bi}(f,\alpha,\alpha)\approx RCS_{Mono}\left(f,\alpha-\frac{\beta}{2},\alpha+\frac{\beta}{2}\right).
\end{equation}
According to a comprehensive investigation \cite{30}, the MBET demonstrates strong performance for simple geometries, maintaining accuracy for Bi-static angles of up to 30° (including sidelobe structure and amplitudes). However, for minimally complex objects, the accuracy of the MBET diminishes significantly, remaining reliable only for Bi-static angles less than 15–20°. Future research is needed to develop modeling methods for Bi-static RCS at larger Bi-static angle $\beta$.

\section{Metrics for ISAC Evaluation}
\subsection{Accuracy metric}

\begin{table}[!b] 
\footnotesize
\caption{Basic metrics for sensing}
\label{tab2}
\tabcolsep 30pt %space between two columns. ÓÃÓÚµ÷ÕûÁÐ¼ä¾à
\renewcommand{\arraystretch}{1.5} 
\begin{tabular*}{\textwidth}{cccc}
\toprule
  Sensing usage & Principle &Accuracy &Resolution \\\hline
  Ranging &$R = \frac{1}{2} c t_{R}$& $d R = \frac{1}{2} c dt_{R} + \frac{R}{C} dc$ & $\Delta R = \frac{1}{2} c t_{\tau}$ \\\hline
  Speed detection & $f_d = \frac{2 v}{\lambda}$&$d v = \frac{\lambda}{2}d f_d$  &$\Delta v = \frac{\lambda}{2} \Delta f_d$ \\\hline
  Angulation & $\theta = \arcsin(\frac{\phi \lambda}{2 \pi d})$ &$d\theta = \frac{\lambda d\phi}{2 \pi d \cos{\theta}}$ & $\Delta\theta > \frac{\lambda}{Nd\cos{\theta}}$ \\
\bottomrule
\end{tabular*}
\end{table}
The fundamental detection equation in radar is presented in Table \ref{tab2}.
Here, $c$ is the speed of light and $t_R$ represents the time interval between the transmitted pulse and the received echo pulse from the sensed target. The ranging accuracy can be qualitatively analyzed using the total differential expression as defined in Table \ref{tab2}.
The range resolution $\Delta R$ is defined as the minimum distinguishable distance, which is related to the pulse width $t_{\tau}$ of the sensing signal.
The maximum unambiguous range is determined by the pulse repetition interval $T_{R}$ and is expressed as follows:
\begin{equation}
\label{eq_MaxRange}
R_{Max} = \frac{1}{2} c T_{R},
\end{equation}

The basic principle of angle measurement in the radar field is as follows
\begin{equation}
\label{eq_ADtheory1}
\phi = \frac{2\pi}{\lambda} d \sin{\theta},
\end{equation}
where $\phi$ represents the phase difference between different antenna elements that are separated by a distance $d$ at the receiver and $\theta$ is the angle of the sensing target. The angle measurement accuracy can be expressed as  
\begin{equation}
\label{eq_ADaccuracy1}
d\phi = \frac{2\pi}{\lambda} d \cos{\theta} d\theta,
\end{equation}

\begin{equation}
\label{eq_ADfuzziness}
\phi = \frac{2 \pi}{\lambda} d \sin{\theta} < \pi.
\end{equation}
Based on equation (\ref{eq_ADfuzziness}), we can drive $\theta_{max}=\arcsin{(\frac{\lambda}{2\pi})}$.
The angle resolution $\Delta \theta$ is defined as the minimum distinguishable angle, which is related to the main lobe width of the antenna array $\frac{2 \pi}{N}$.
The basic principle of velocity measurement is based on the Doppler frequency.
From Table \ref{tab2}, we obtain $v=\frac{\lambda f_d}{2}$. Analogous to range resolution , angular resolution and Doppler resolution can be expressed as $\Delta f_d = \frac{1}{M T_R}$, in which $M$ is the number of transmitted pulses, and $T_R$ is the pulse repetition interval.

\subsection{Detection metric}
The probability of false alarms arises when the noise at the receiver surpasses the decision threshold, resulting in erroneous detection. This probability is associated with the root mean square (RMS) value of the noise and the decision threshold, and it is defined as follows
\begin{equation}
\label{eq_Pfa1}
P_{fa} = \int_{V_T}^{\infty} \omega(x|\text{no target}) d x,
\end{equation}
where $V_T$ is the decision threshold, $\omega(x)$ denotes the probability density function (PDF) of detection. When the noise is complex and both its real and imaginary components follow a Gaussian distribution, the magnitude of the noise follows a Rayleigh distribution, $\omega(x)$ can be expressed as 
\begin{equation}
\label{eq_Pfaw}
\omega(x|\text{no target}) = \frac{x}{\sigma^2} exp(-\frac{x^2}{2\sigma^2}),x\geq0. 
\end{equation}
Here, $\sigma$ represents the standard deviation of the noise. Substituting equation (\ref{eq_Pfaw}) into equation (\ref{eq_Pfa1}) results in
\begin{equation}
\label{eq_Pfa2}
P_{fa} = exp(-\frac{V_T^2}{2 \sigma^2}).
\end{equation}
Further, the relationship between $P_{fa}$ and $V_T$ is obtained as $ V_T = \sqrt{-2\ln{P_{fa}}}\sigma$. The probability of detection \( P_D \) represents the likelihood of correctly detecting a target when one is present and is defined as follows
\begin{equation}
\label{eq_Pd}
P_{d} = \int_{V_T}^{\infty} \omega(x|\text{has target}) d x = \int_{V_T}^{\infty} \frac{x}{\sigma} exp(-\frac{A^2+x^2}{2\sigma^2}) J_0(\frac{Ax}{\sigma^2}) dx.
\end{equation}

\begin{figure}[!b]
	\centering
	\begin{minipage}{0.46\linewidth}
		\centering
		\includegraphics[width=1\linewidth]{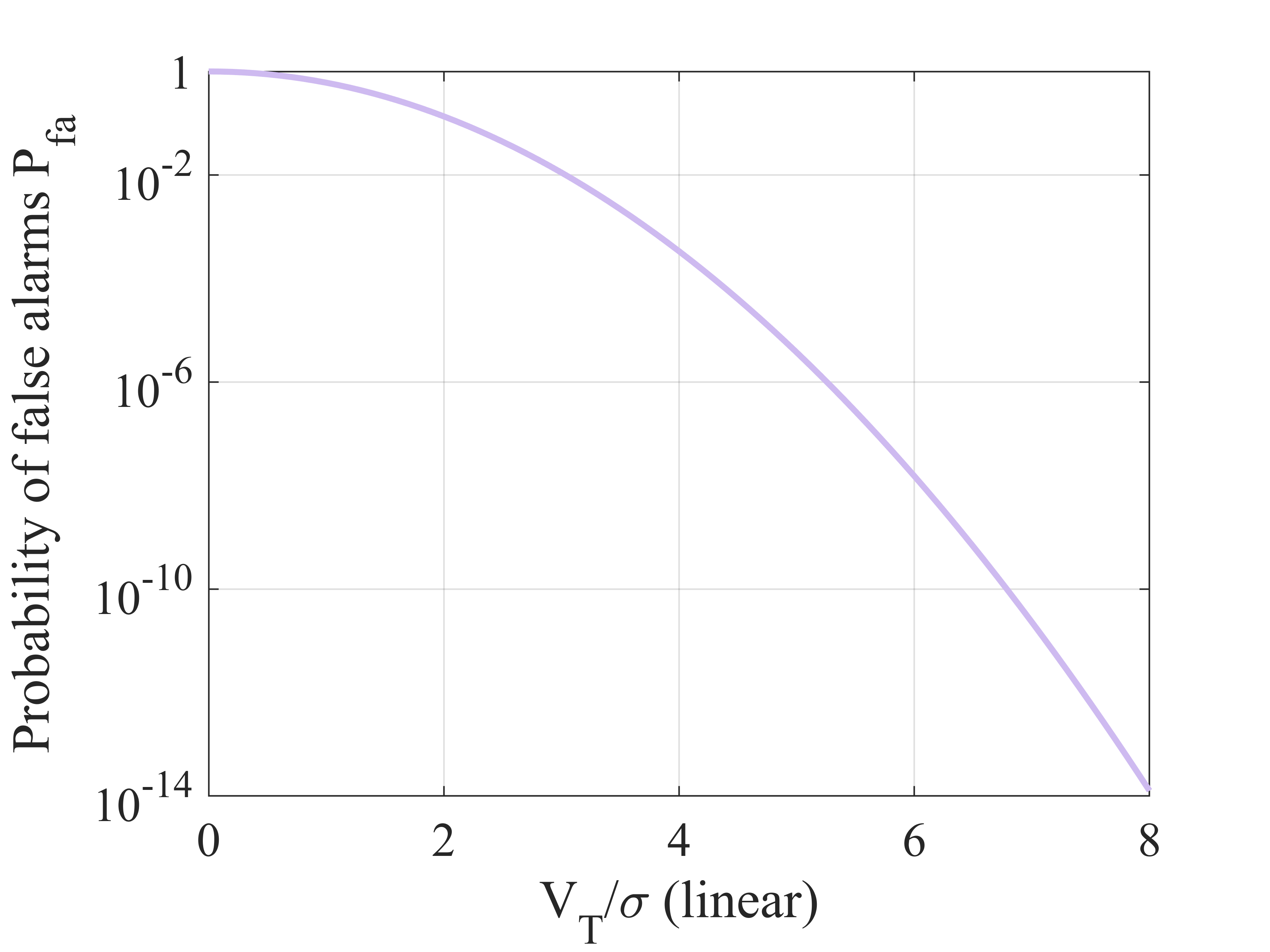}
		\caption*{(a)}
		%\label{fig_ASA}
	\end{minipage}
	\begin{minipage}{0.46\linewidth}
		\centering
		\includegraphics[width=1\linewidth]{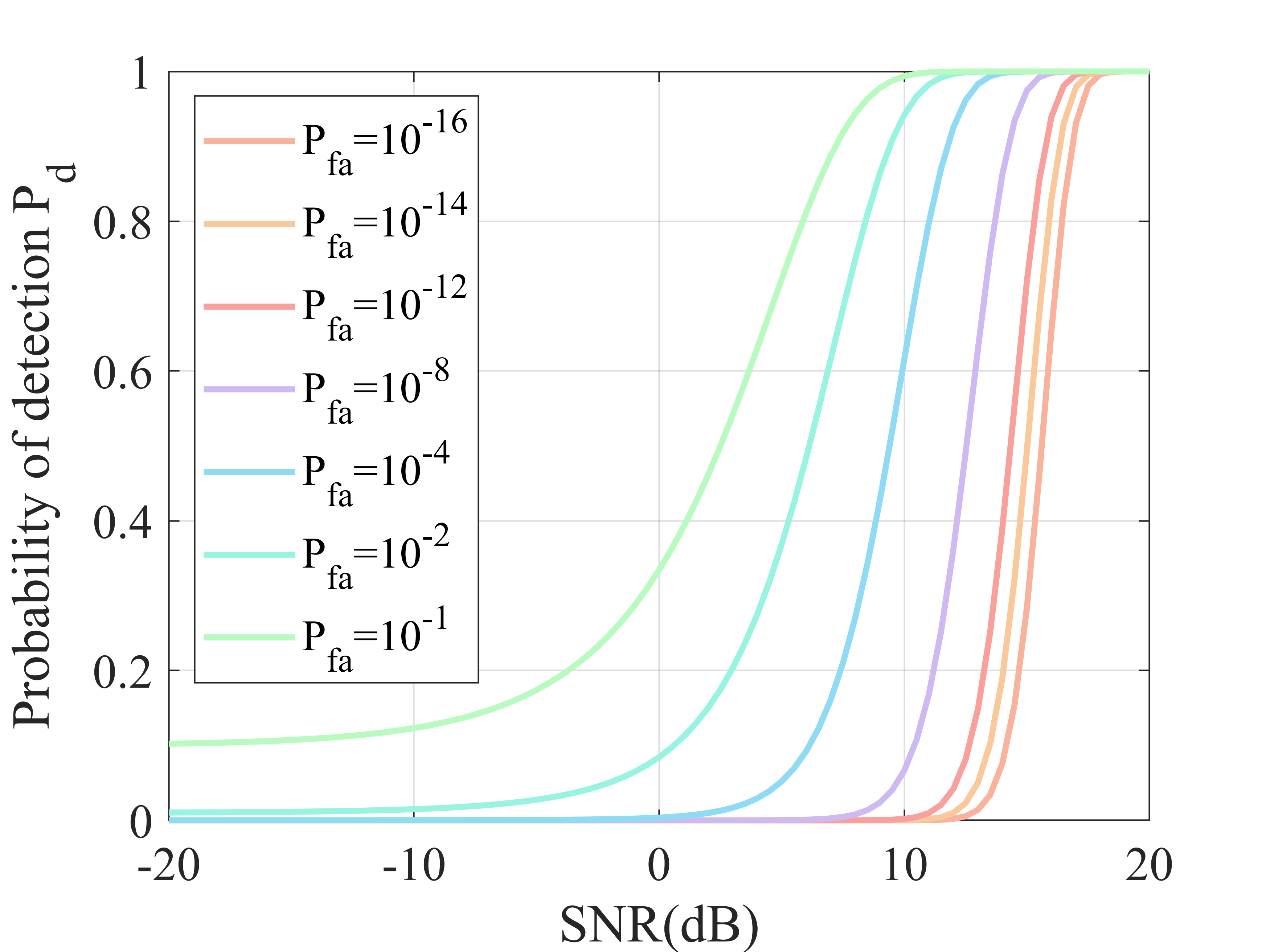}
		\caption*{(b)}
		%\label{fig_ASD}
	\end{minipage} 
    \caption{Relationship between detection metrics and system performance. (a) $P_{fa}$ versus the linear value of $V_T/\sigma$. (b) $P_d$ versus SNR (dB) under different $P_{fa}$ conditions.}
 \label{Pd}
\end{figure}
Here, $A$ denotes the signal amplitude, and $J_0(\cdot)$ represents the modified Bessel function of the first kind. Further exploration of the relationship between the proposed sensing performance metrics and traditional system performance metrics, such as bit error rate (BER), symbol error rate (SER), and signal-to-noise ratio (SNR), holds significant reference value for ISAC system design and testing. Considering that both SER and BER are also related to SNR, the aforementioned accuracy metrics are directly influenced by the parameters used in the detection process such as ranging and delay, velocity and Doppler, and angle and phase. Therefore, we will focus on investigating the interplay between SNR and detection metrics (probability of detection and false alarm probability).

As shown in Figure \ref{Pd}(a), the false alarm probability decreases exponentially with respect to the threshold level. When the noise distribution function is fixed, the magnitude of the false alarm probability is entirely dependent on the threshold level, which decreases exponentially as the threshold increases. As shown in Figure \ref{Pd}(b), for a fixed false alarm probability, the detection probability increases with increasing SNR, i.e., for a constant threshold level, the detection probability increases as the SNR increases. In other words, for a fixed SNR, a lower false alarm probability corresponds to a lower detection probability, and a higher false alarm probability corresponds to a higher detection probability. By combining Figure \ref{Pd}(a) and \ref{Pd}(b), it is evident that when the relative threshold $V_{T}/\sigma$ increases, the false alarm probability decreases, but the detection probability also decreases. We always aim to improve the detection probability for a fixed false alarm probability, which can only be achieved by increasing the SNR.

\section{Simulation Results}
\subsection{Concatenation methods}
In the simulation experiments, the primary focus is on the reliability of the ISAC channel concatenated method and the accuracy of the RCS simulation results. RAN1 \#118bis meeting has agreed 8 options for concatenating the Tx-target and target-Rx link in the target channel as
\begin{itemize}
    \item Direct path (if present) is always kept
    \item Indirect paths of LOS+NLOS, NLOS+LOS (if present) are generated
    \item On other indirect paths of NLOS + NLOS 
    \begin{itemize}
        \item[-]  Option 0: ray level full convolution between Tx-target and target-Rx link for Case 1/2/3/4.
        \item[-] Option 0A: ray level full convolution between Tx-target and target-Rx link only for Case 4.
        \item[-] Option 1: cluster level full convolution between Tx-target link and target-Rx link, then 1-by-1 coupling rays within each pair of clusters for Case 1/2/3/4.
        \item[-] Option 1A: cluster level full convolution between Tx-target and target-Rx link, then 1-by-1 coupling rays within each pair of clusters only for Case 4.
        \item[-] Option 2: cluster level 1-by-1 coupling between Tx-target and target-Rx link, then 1-by-1 coupling rays within each pair of clusters for Case 1/2/3/4.
        \item[-] Option 2A: cluster level 1-by-1 coupling between Tx-target and target-Rx link, then 1-by-1 coupling rays within each pair of clusters only for Case 4.
        \item[-] Option 3: ray level 1-by-1 coupling between Tx-target and target-Rx link for Case 1/2/3/4.
        \item[-] Option 3A: ray level 1-by-1 coupling between Tx-target and target-Rx link only for Case 4.
    \end{itemize}
\end{itemize}

Here, Case 1, 2, 3, and 4 represent link conditions such as LOS+LOS, LOS+NLOS, NLOS+LOS, and NLOS+NLOS respectively. Metrics defined for the target channel include power, delay spread (DS), and angle spread (ASA, ASD, ZSA, ZSD). Based on the concatenation methods at the cluster level, ray level, random coupling, and order coupling, the 8 options above can be further subdivided into 10 cases:
\begin{table}[!h] 
\footnotesize
\caption{Simulation cases for concatenation}
\label{tab3}
\tabcolsep 10pt 
\renewcommand{\arraystretch}{1.5} 
\begin{tabular*}{\textwidth}{cc}
\toprule

    CaseA & Concatenation only for NLOS+NLOS propagation condition\\
    Case0 & Full convolution between Tx-target link and target-Rx link \\
    Case1 & Cluster full convolution, then 1-by-1 coupling rays within each pair of clusters \\
    Case2O & Cluster level 1-by-1 coupling, then 1-by-1 coupling  rays in order within each pair of clusters \\
    Case2R & Cluster level 1-by-1 coupling, then 1-by-1 randomly coupling rays within each pair of clusters \\
    Case3 & Ray level 1-by-1 randomly coupling between Tx-target link and target-Rx link \\
    Case1N & Cluster full convolution, 1-by-1 coupling rays (NLOS-NLOS normalization) \\
    Case2ON & Cluster level 1-by-1 coupling, 1-by-1 coupling rays in order  (NLOS-NLOS normalization) \\
    Case2RN & Cluster level 1-by-1 coupling, then 1-by-1 randomly coupling rays  (NLOS-NLOS normalization) \\
    Case3N & Ray level 1-by-1 randomly coupling between Tx-target link and target-Rx link (NLOS-NLOS normalization) \\
\bottomrule
\end{tabular*}
\end{table}
Options 0, 1, 2, and 3A in \#118-bis simulation options for concatenation are uniformly represented by CaseA. The ray-level 1-by-1 coupling in order is excluded, as it is equivalent to Case 2O. Additionally, the specific coupling method of ray 1-by-1, whether sequential or random, does not affect Case 1, Case 2O, or Case 2R. This is because the power of each ray pair depends solely on the power of the corresponding clusters, with rays within the same cluster having equal power. Therefore, the coupling method at the ray level does not affect the distribution, as the power of each ray pair remains the same once the cluster-level concatenation is complete. The simulation configuration parameters and corresponding results are presented in Table \ref{tab4}.
\begin{table}[!t] 
\footnotesize
\caption{Simulation setups}
\label{tab4}
\tabcolsep 80pt 
\renewcommand{\arraystretch}{1.5} 
\begin{tabular*}{\textwidth}{cc}
\toprule

  Scenarios & UMi in 3GPP TR38.901  \\
  Sensing mode  & Bi-static \\
  Frequency & $6GHz$ \\
  RCS model & $RCS=A*B_1*B_2=1$ \\
\bottomrule
\end{tabular*}
\end{table}

\begin{figure}[!b]
	\centering
	\begin{minipage}{0.245\linewidth}
		\centering
		\includegraphics[width=1\linewidth]{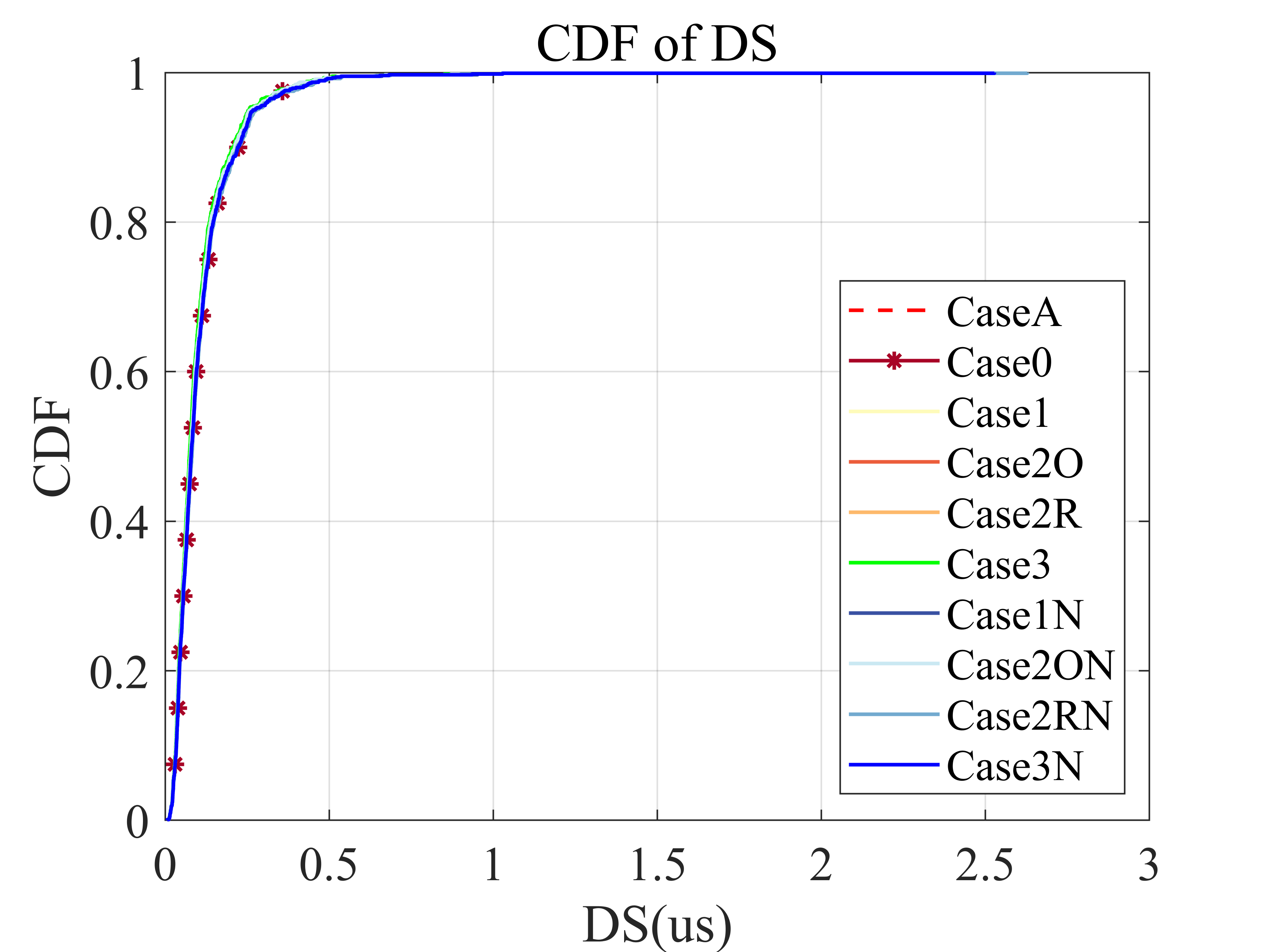}
		\caption*{DS of LOS case 1}
	
	\end{minipage}
	\begin{minipage}{0.245\linewidth}
		\centering
		\includegraphics[width=1\linewidth]{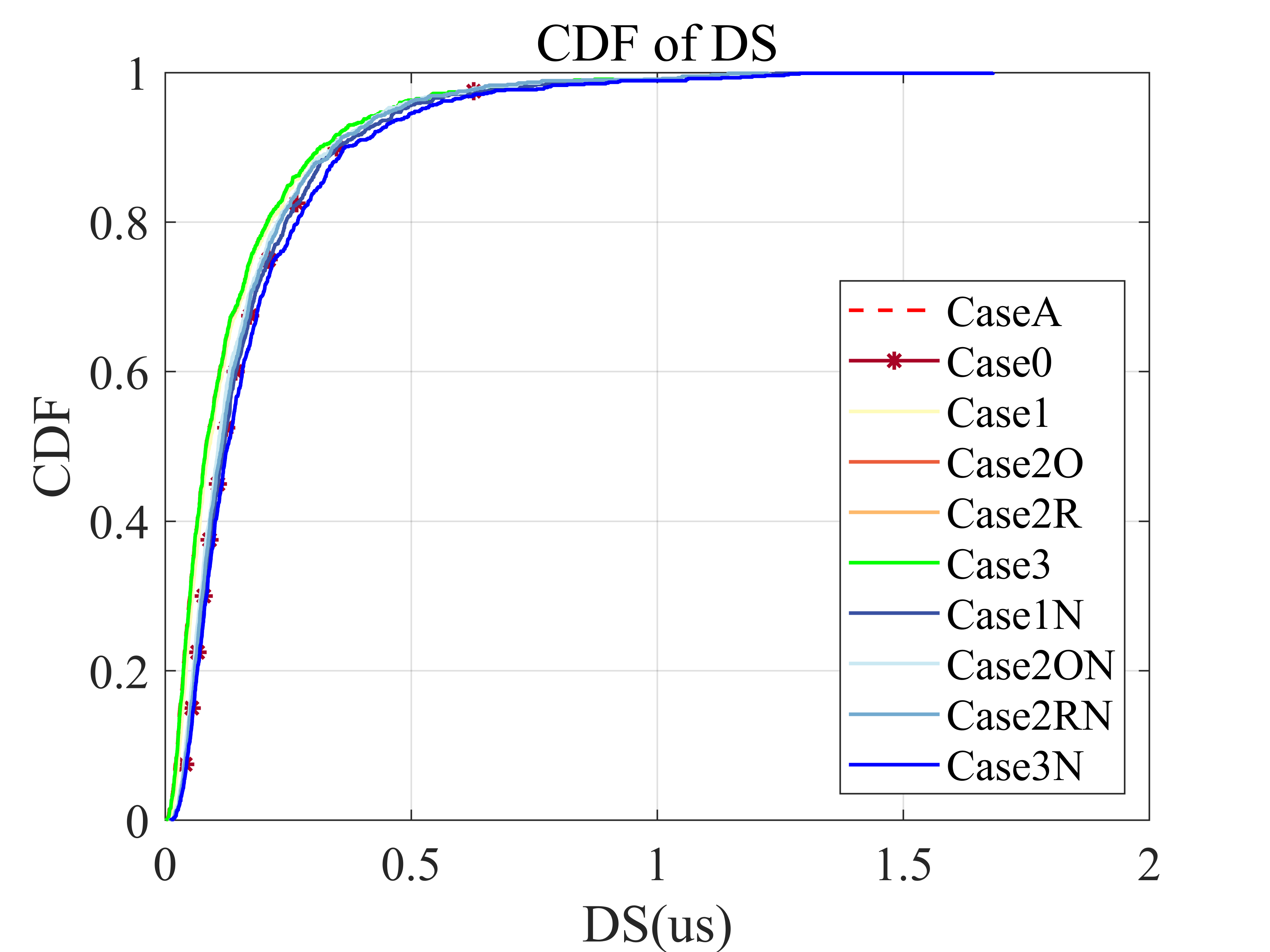}
		\caption*{DS of LOS case 2}
	
	\end{minipage}
    \begin{minipage}{0.245\linewidth}
		\centering
		\includegraphics[width=1\linewidth]{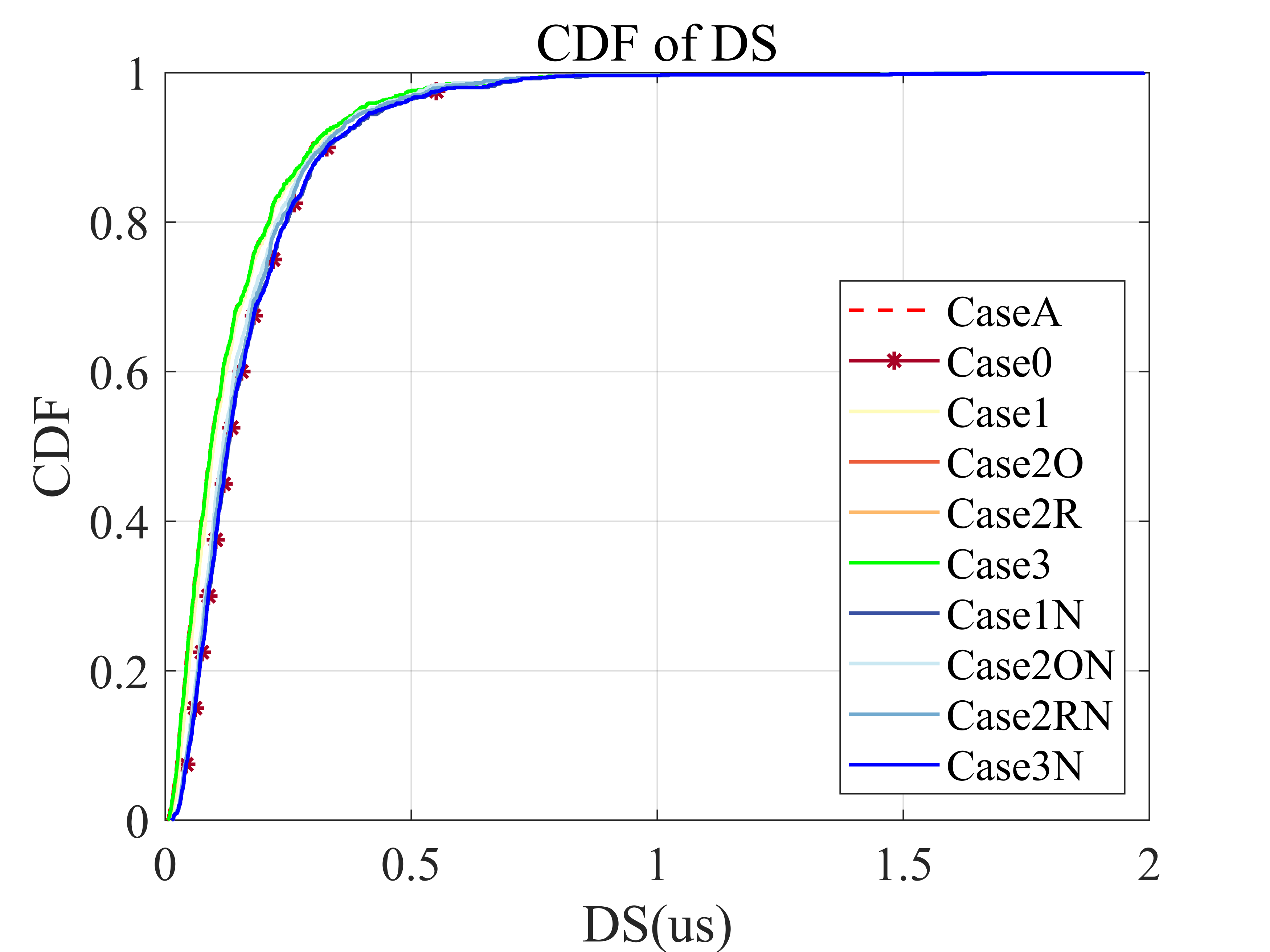}
		\caption*{DS of LOS case 3}
		
	\end{minipage}
    \begin{minipage}{0.245\linewidth}
		\centering
		\includegraphics[width=1\linewidth]{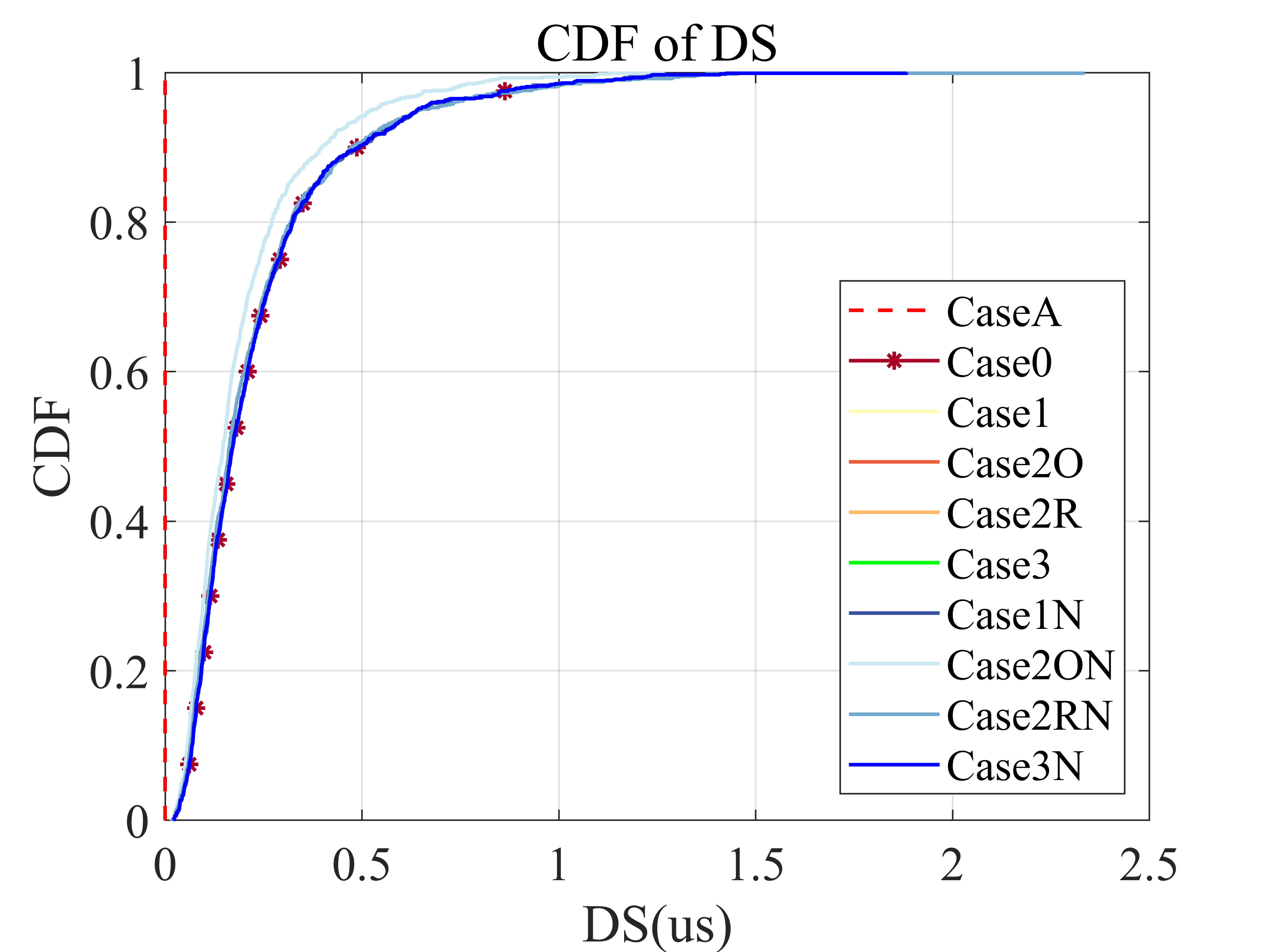}
		\caption*{DS of LOS case 4}
	
	\end{minipage}
	\\

	\begin{minipage}{0.245\linewidth}
		\centering
		\includegraphics[width=1\linewidth]{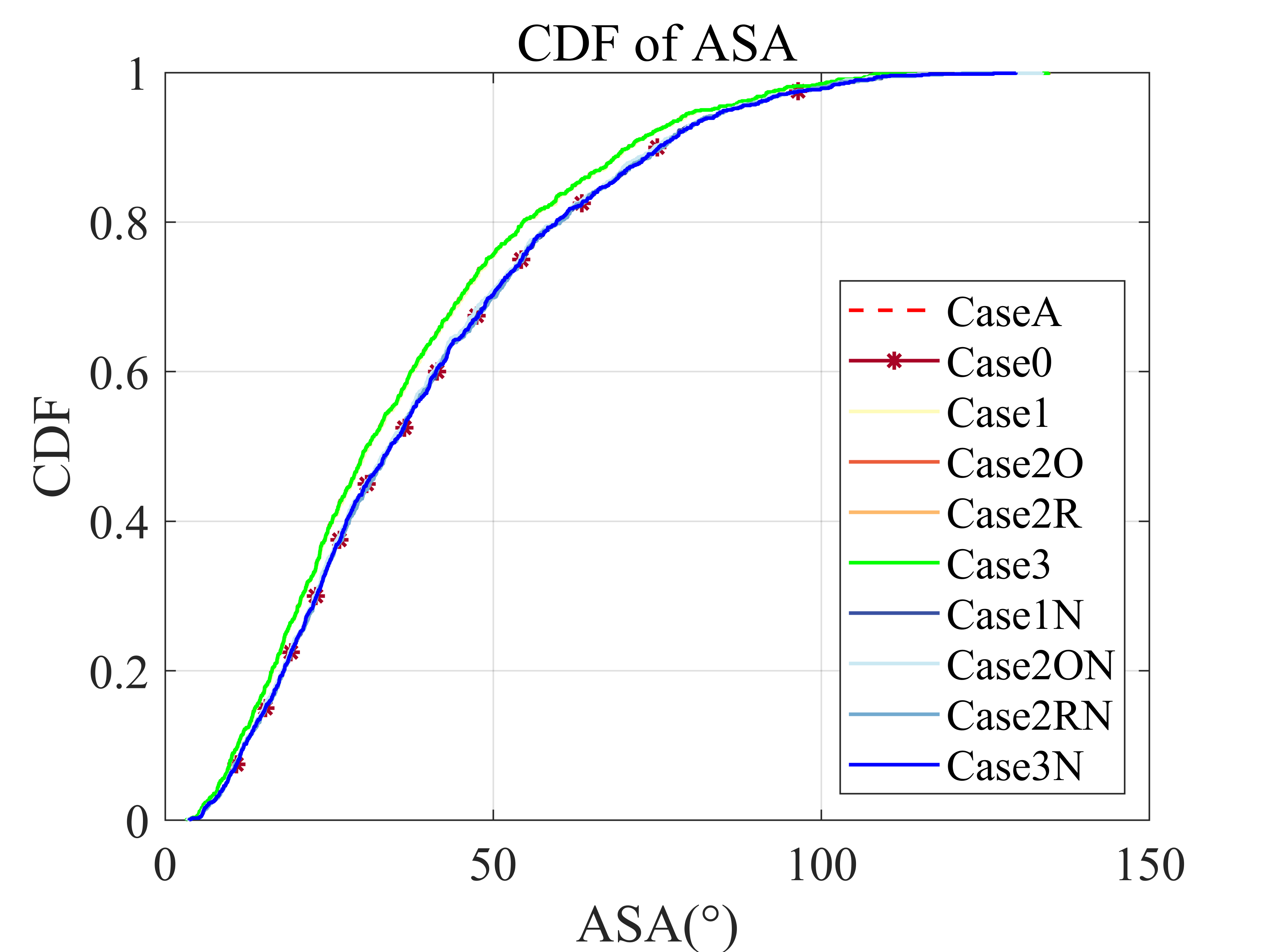}
		\caption*{ASA of LOS case 1}
	
	\end{minipage}
	\begin{minipage}{0.245\linewidth}
		\centering
		\includegraphics[width=1\linewidth]{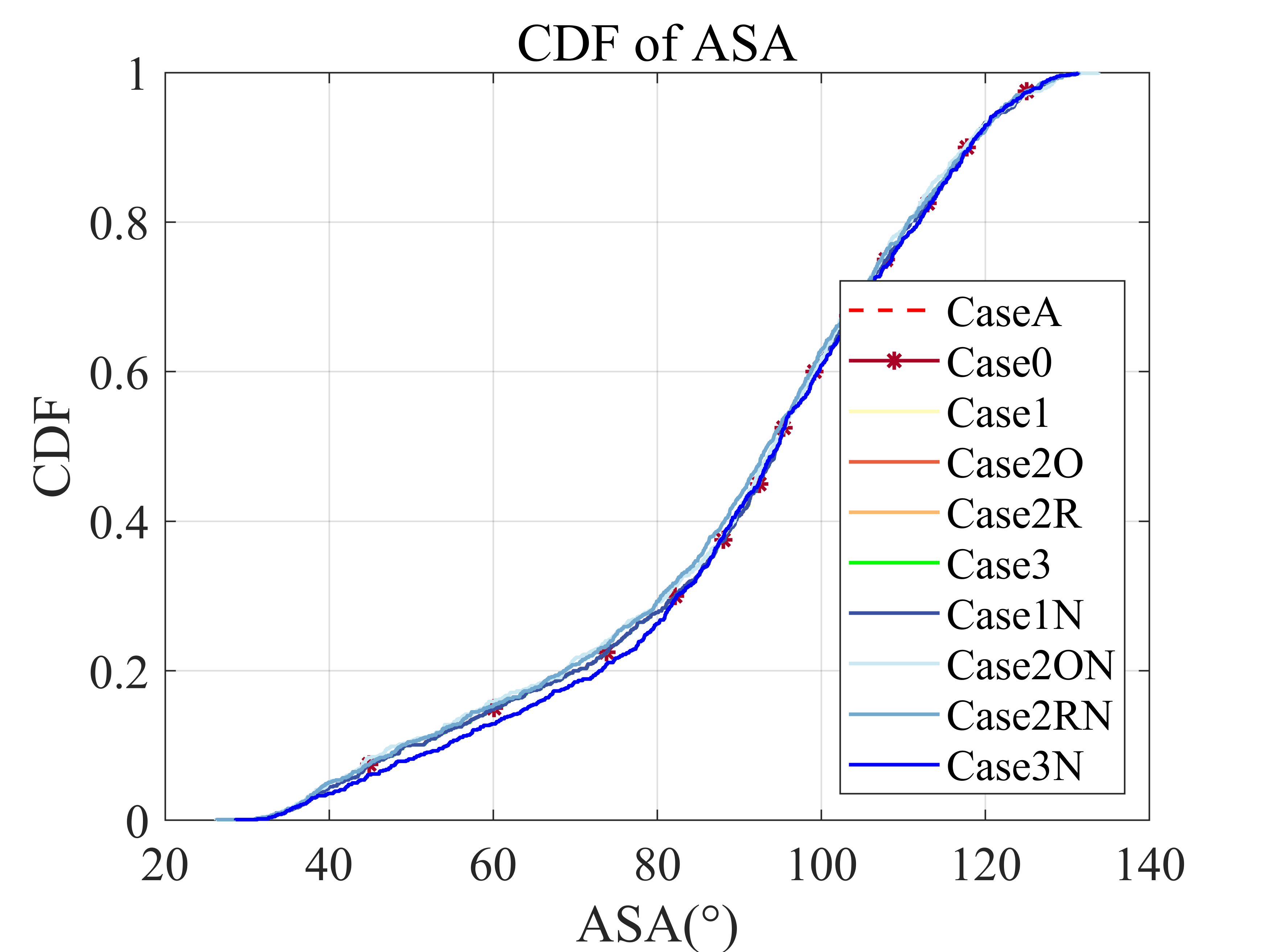}
		\caption*{ASA of LOS case 2}
		
	\end{minipage}
    \begin{minipage}{0.245\linewidth}
		\centering
		\includegraphics[width=1\linewidth]{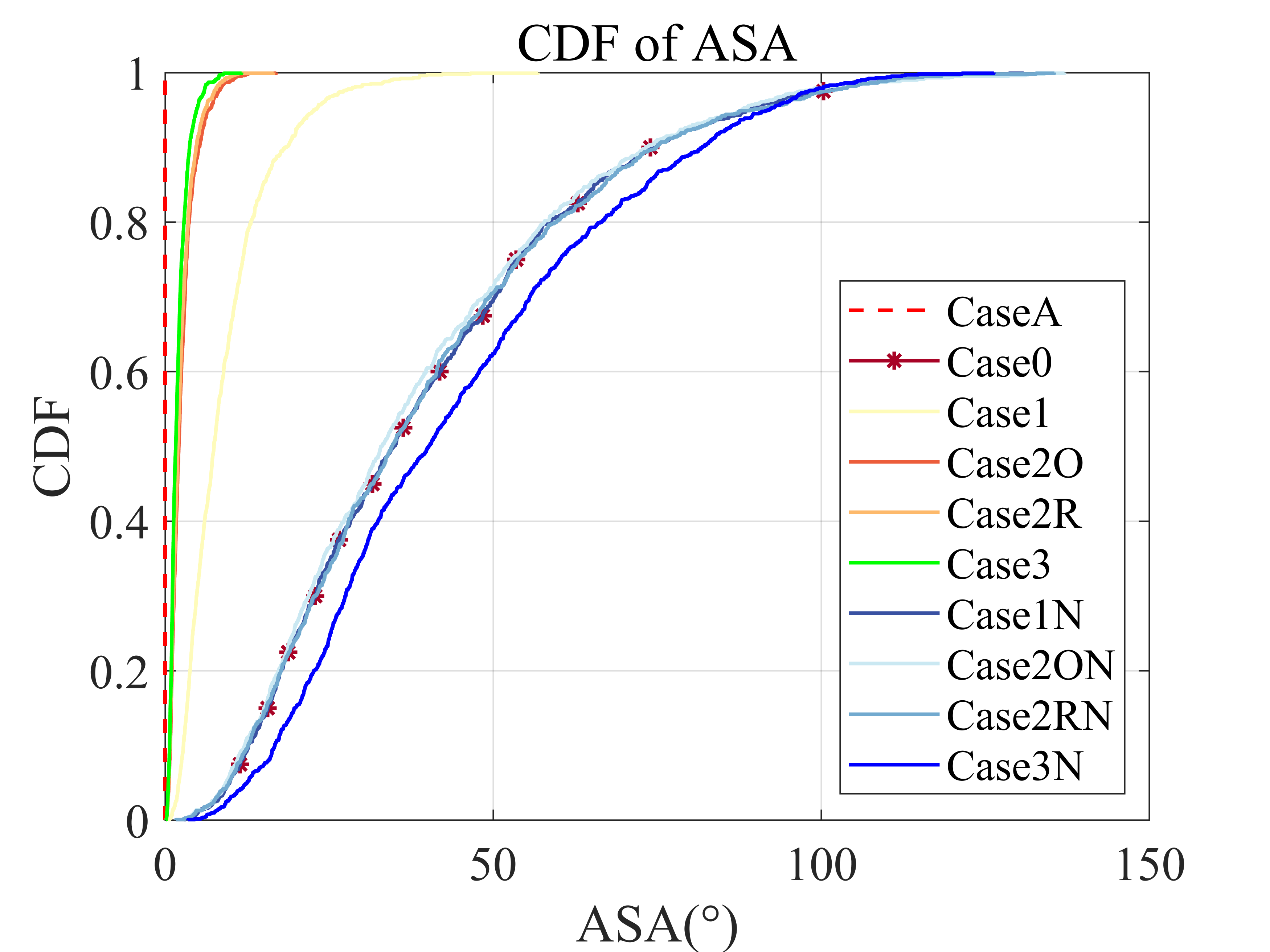}
		\caption*{ASA of LOS case 3}
		
	\end{minipage}
    \begin{minipage}{0.245\linewidth}
		\centering
		\includegraphics[width=1\linewidth]{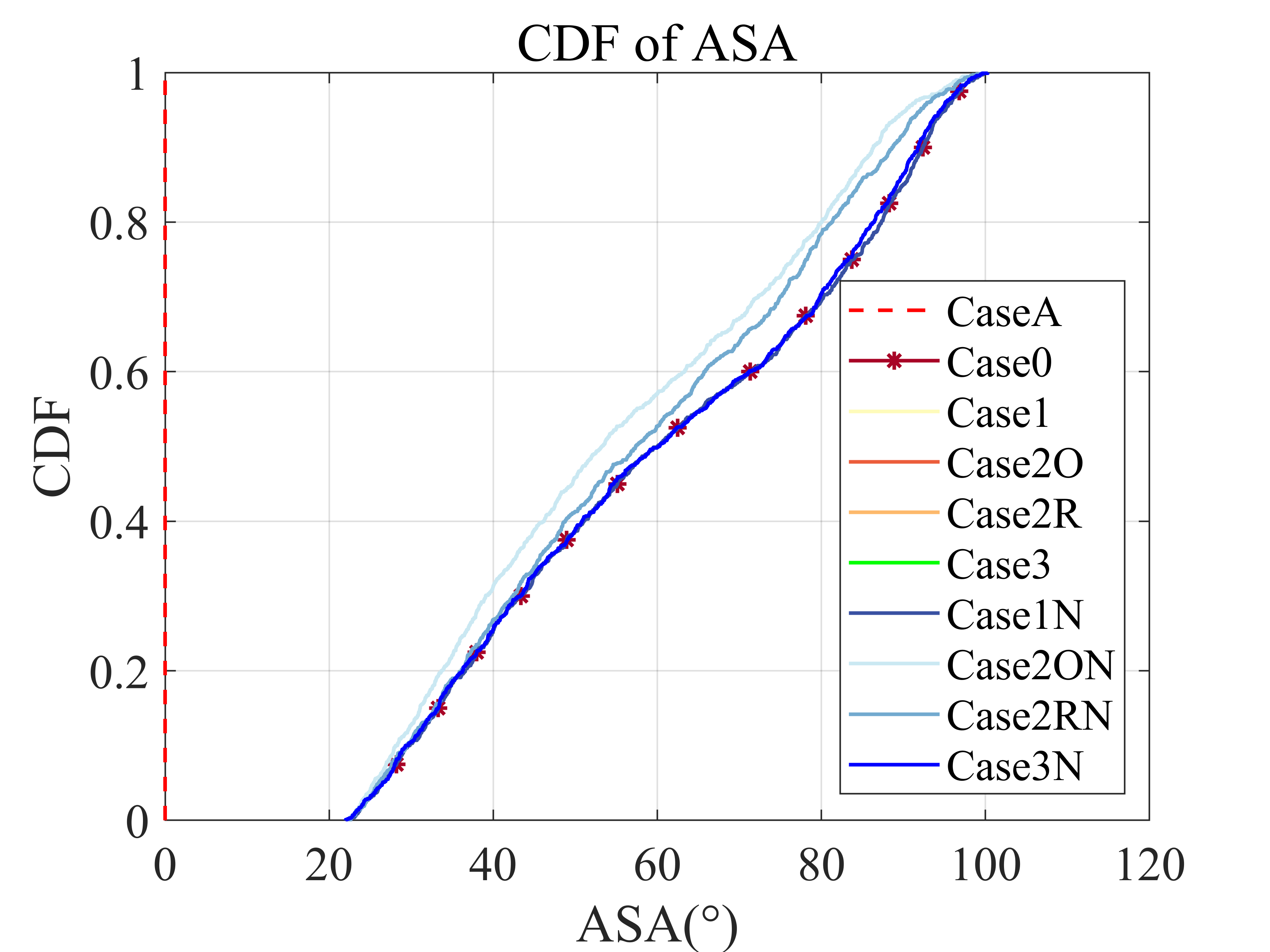}
		\caption*{ASA of LOS case 4}
	
	\end{minipage}
  \\
  	\begin{minipage}{0.245\linewidth}
		\centering
		\includegraphics[width=1\linewidth]{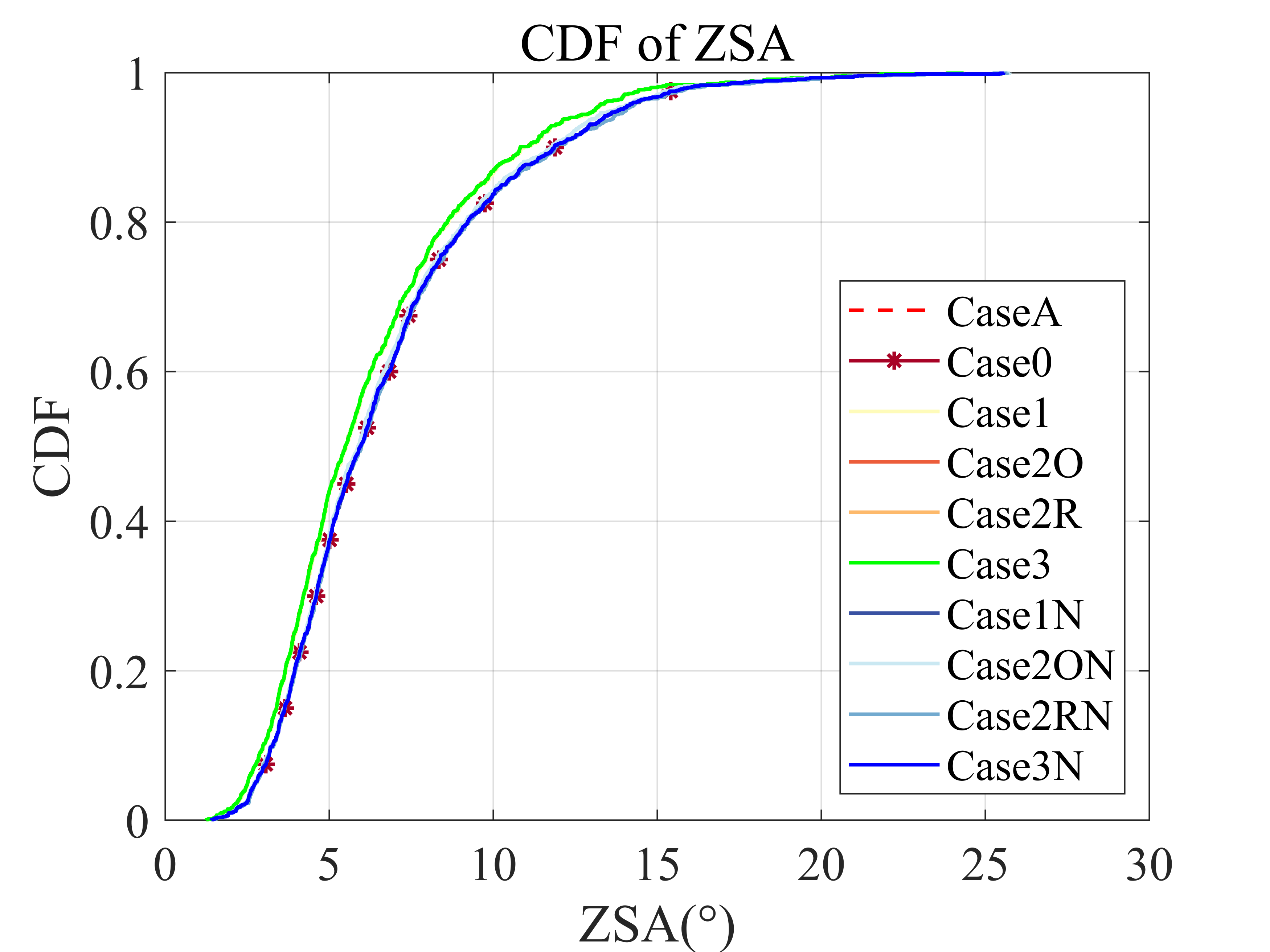}
		\caption*{ZSA of LOS case 1}
		
	\end{minipage}
	\begin{minipage}{0.245\linewidth}
		\centering
		\includegraphics[width=1\linewidth]{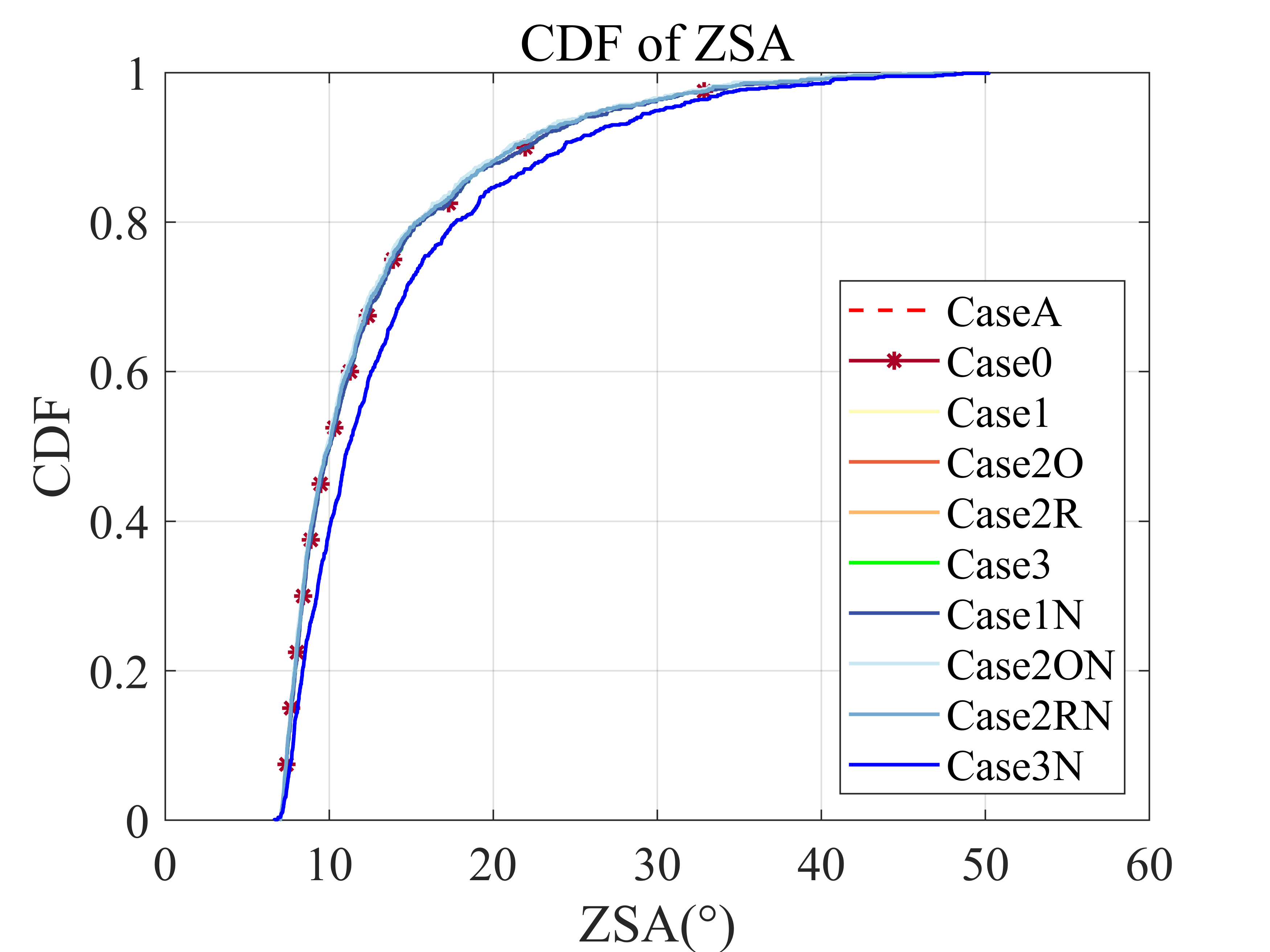}
		\caption*{ZSA of LOS case 2}
		
	\end{minipage}
    \begin{minipage}{0.245\linewidth}
		\centering
		\includegraphics[width=1\linewidth]{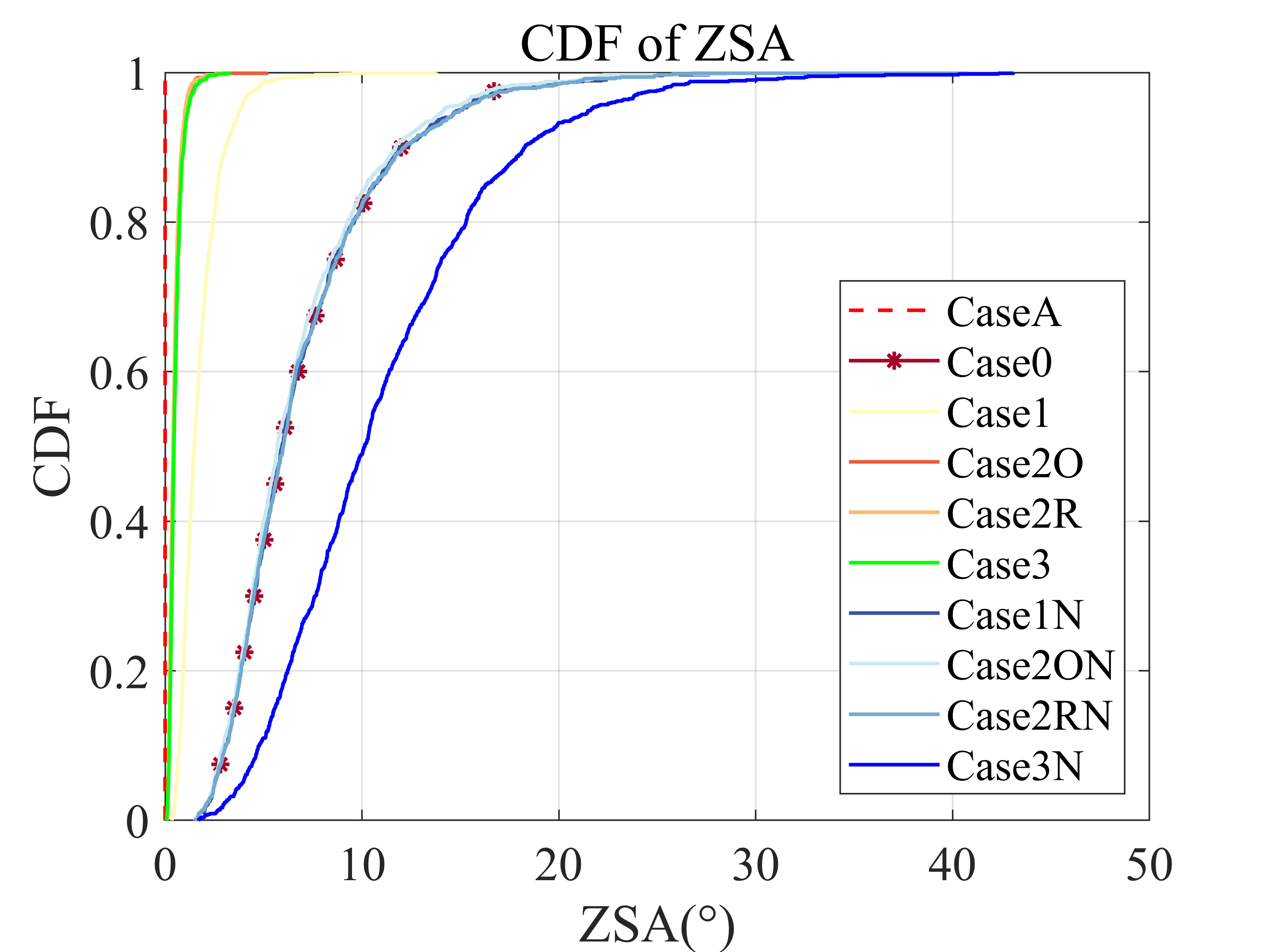}
		\caption*{ZSA of LOS case 3}
	
	\end{minipage}
    \begin{minipage}{0.245\linewidth}
		\centering
		\includegraphics[width=1\linewidth]{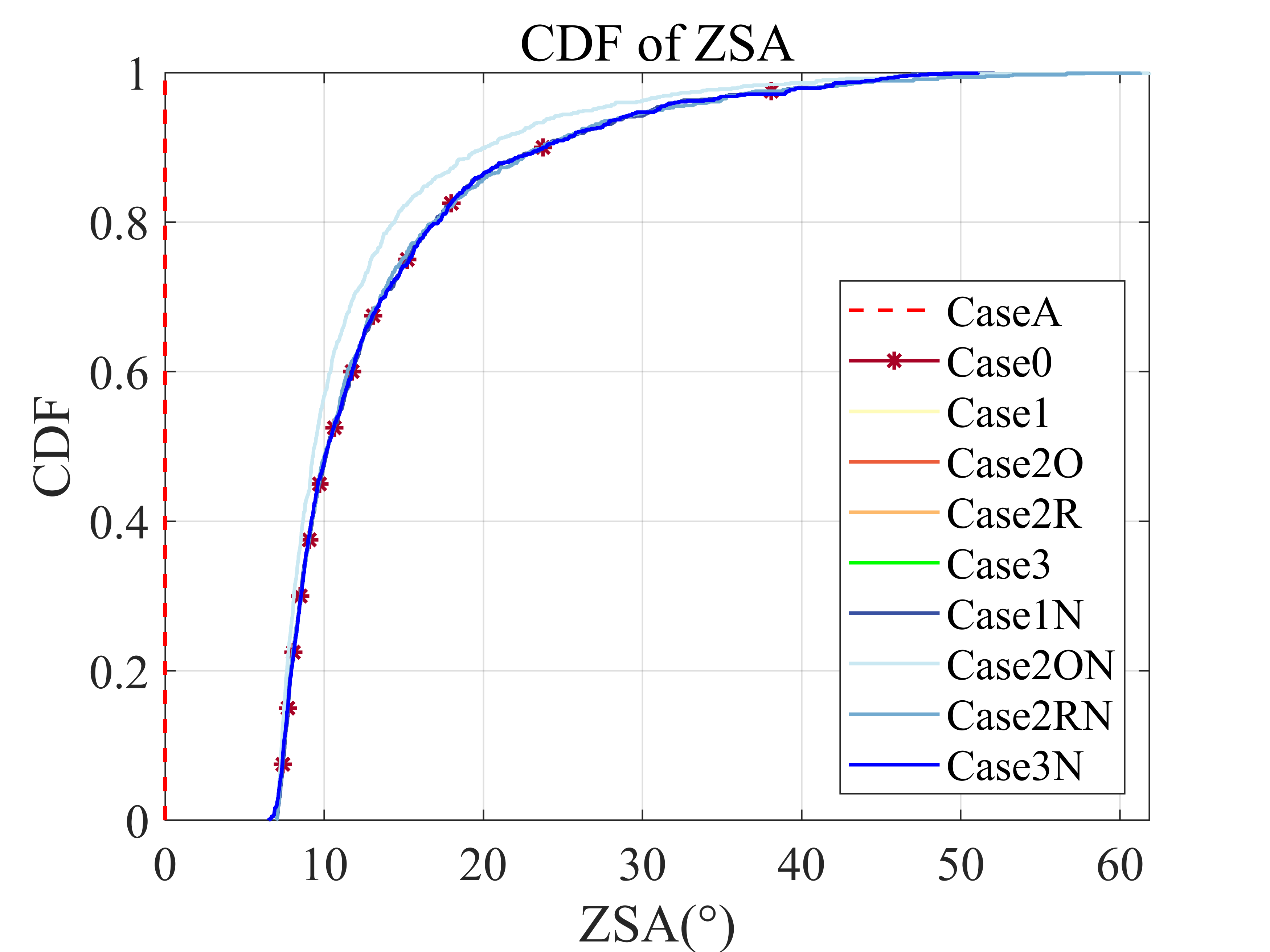}
		\caption*{ZSA of LOS case 4}
	
	\end{minipage}
	\\
    	\begin{minipage}{0.245\linewidth}
		\centering
		\includegraphics[width=1\linewidth]{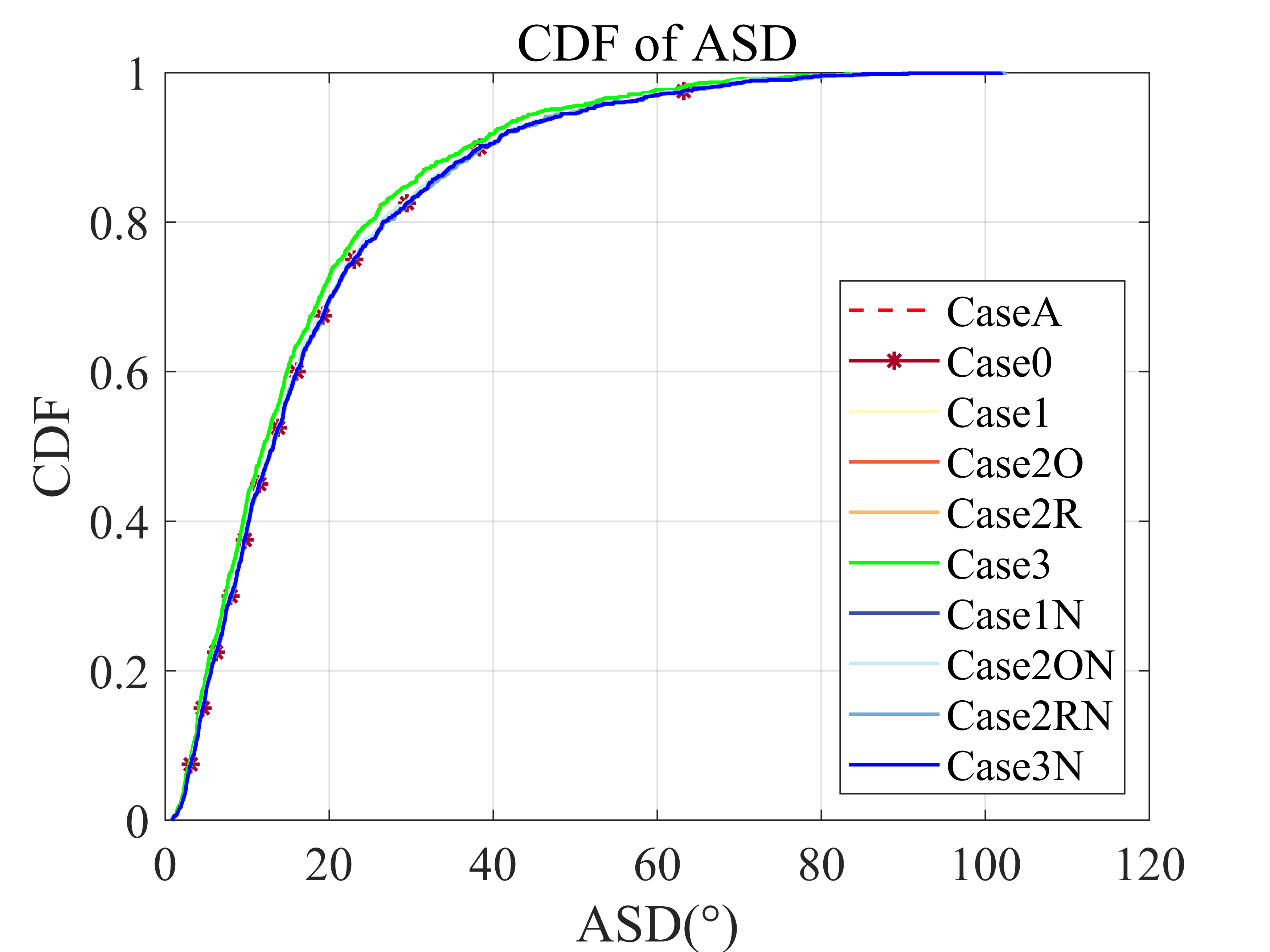}
		\caption*{ASD of LOS case 1}
		
	\end{minipage}
	\begin{minipage}{0.245\linewidth}
		\centering
		\includegraphics[width=1\linewidth]{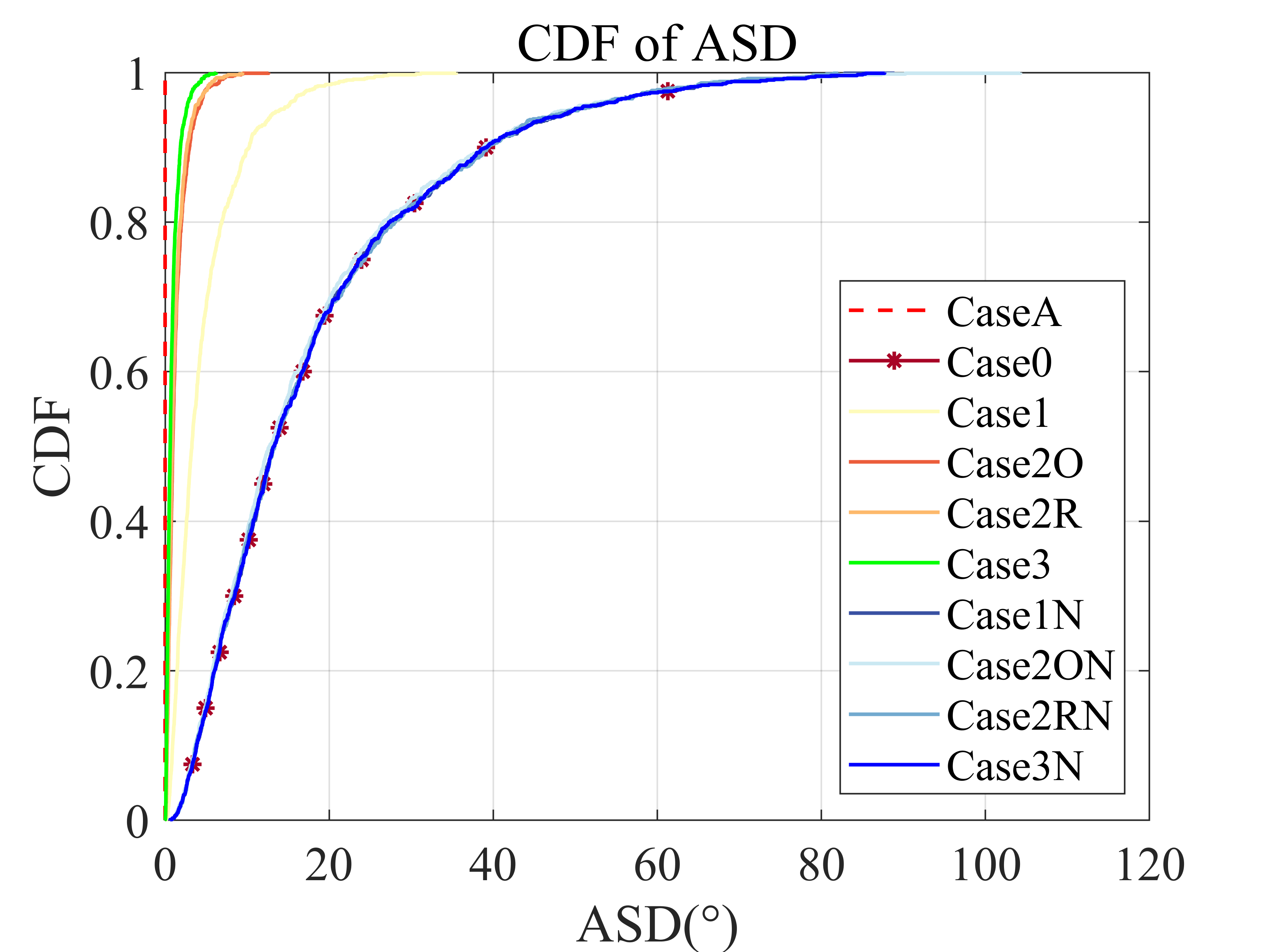}
		\caption*{ASD of LOS case 2}
	
	\end{minipage}
    \begin{minipage}{0.245\linewidth}
		\centering
		\includegraphics[width=1\linewidth]{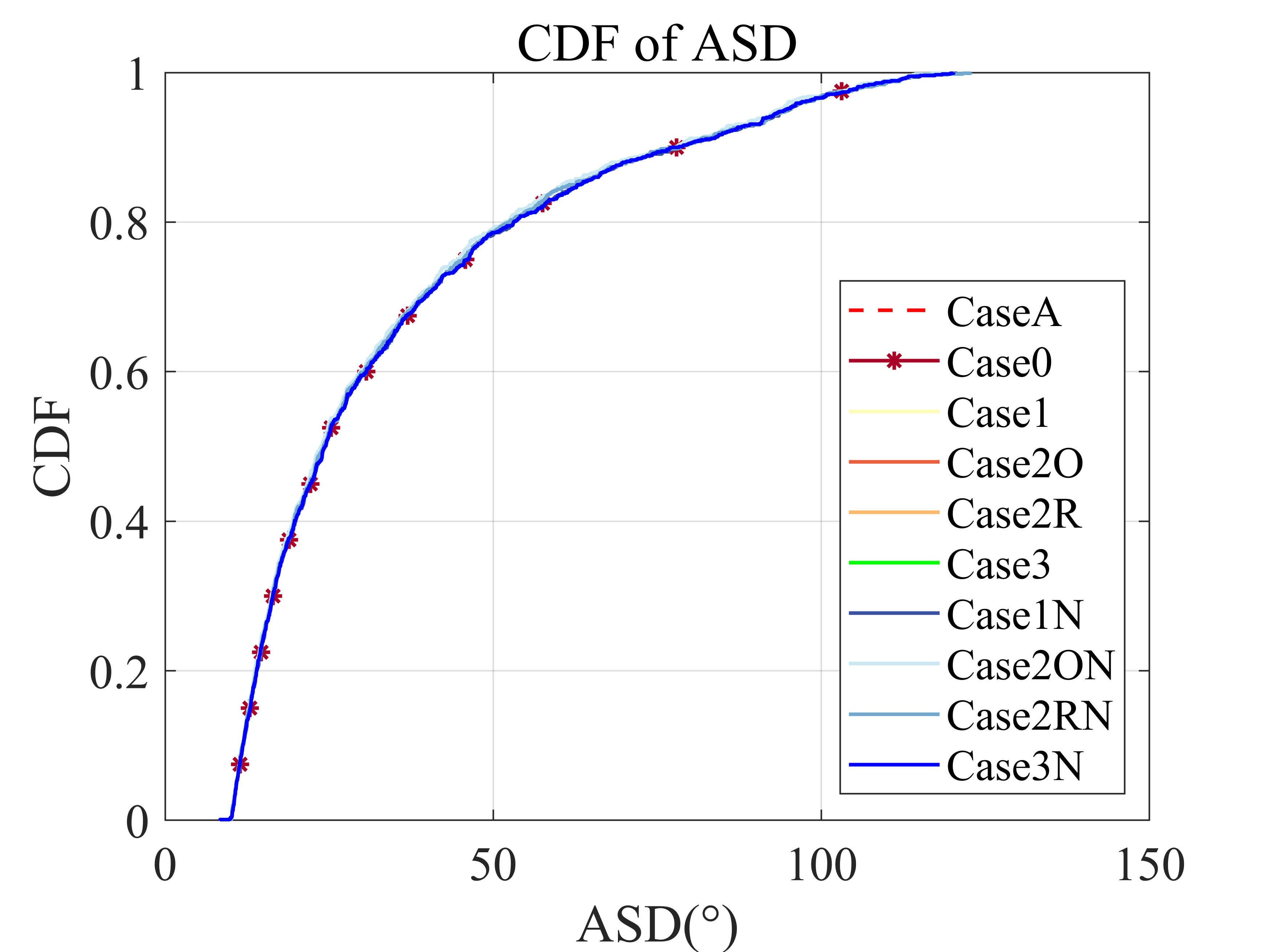}
		\caption*{ASD of LOS case 3}
	
	\end{minipage}
    \begin{minipage}{0.245\linewidth}
		\centering
		\includegraphics[width=1\linewidth]{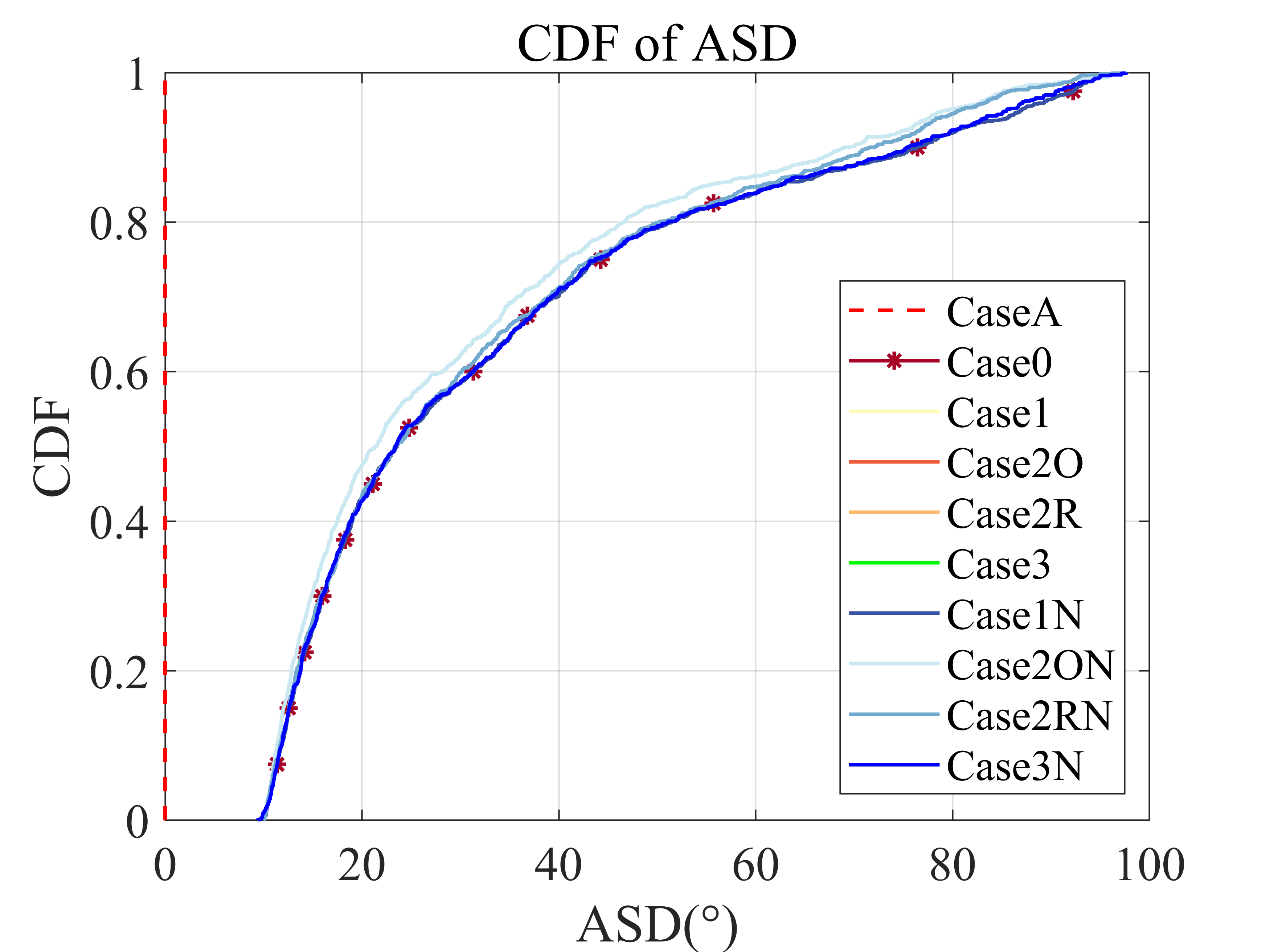}
		\caption*{ASD of LOS case 4}
	
	\end{minipage}
	\\
    	\begin{minipage}{0.245\linewidth}
		\centering
		\includegraphics[width=1\linewidth]{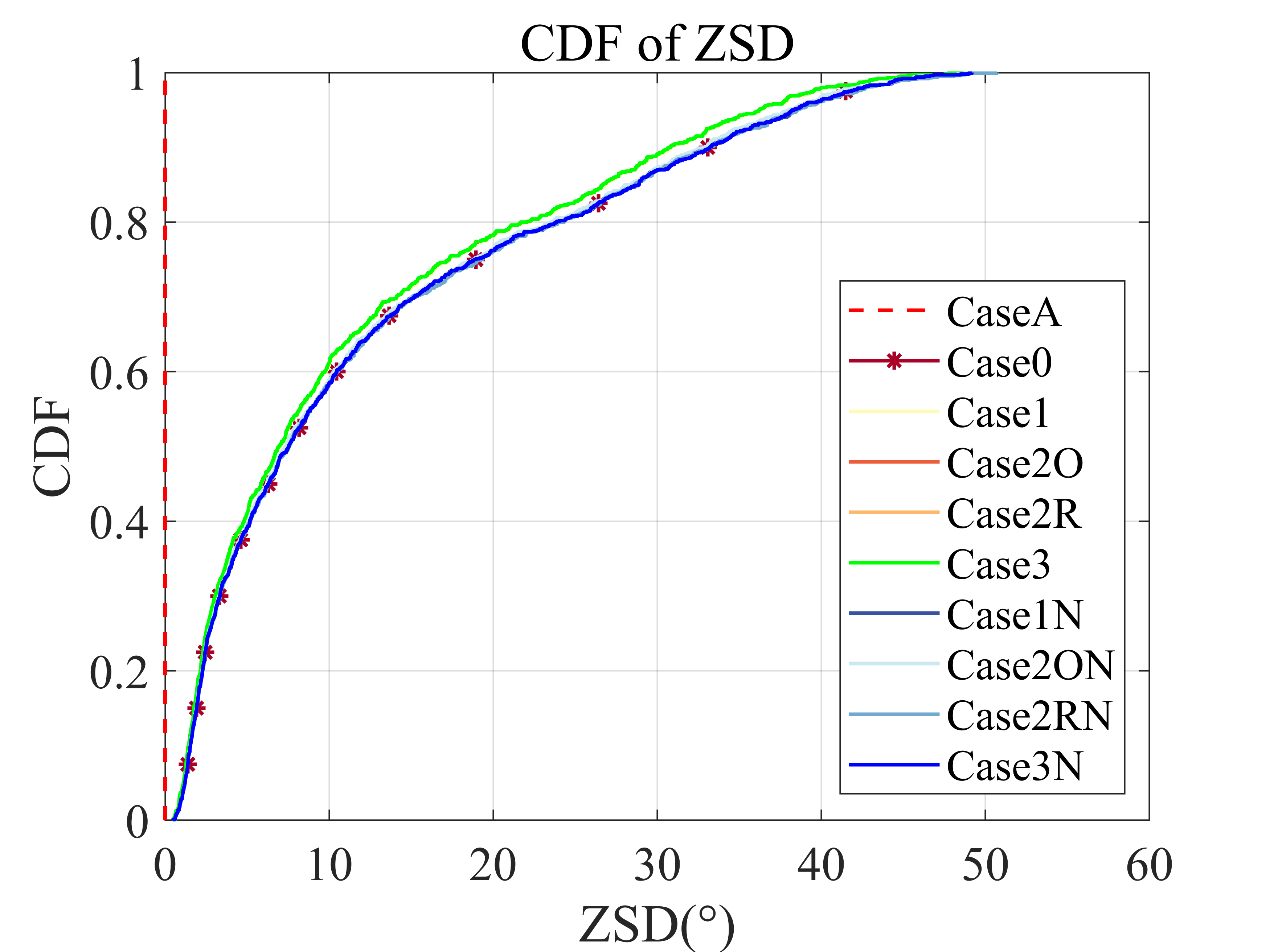}
		\caption*{ZSD of LOS case 1}
	
	\end{minipage}
	\begin{minipage}{0.245\linewidth}
		\centering
		\includegraphics[width=1\linewidth]{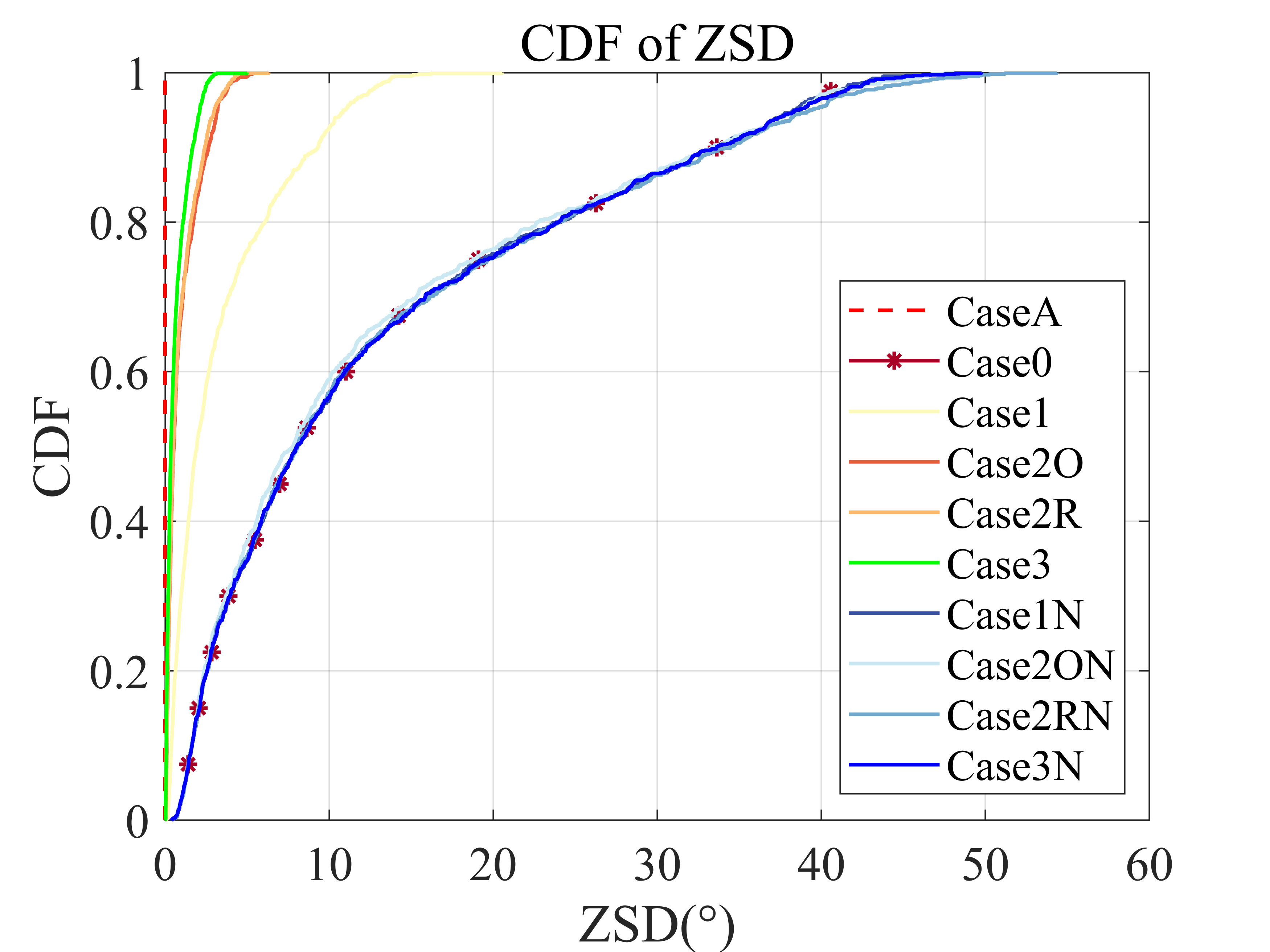}
		\caption*{ZSD of LOS case 2}
	
	\end{minipage}
    \begin{minipage}{0.245\linewidth}
		\centering
		\includegraphics[width=1\linewidth]{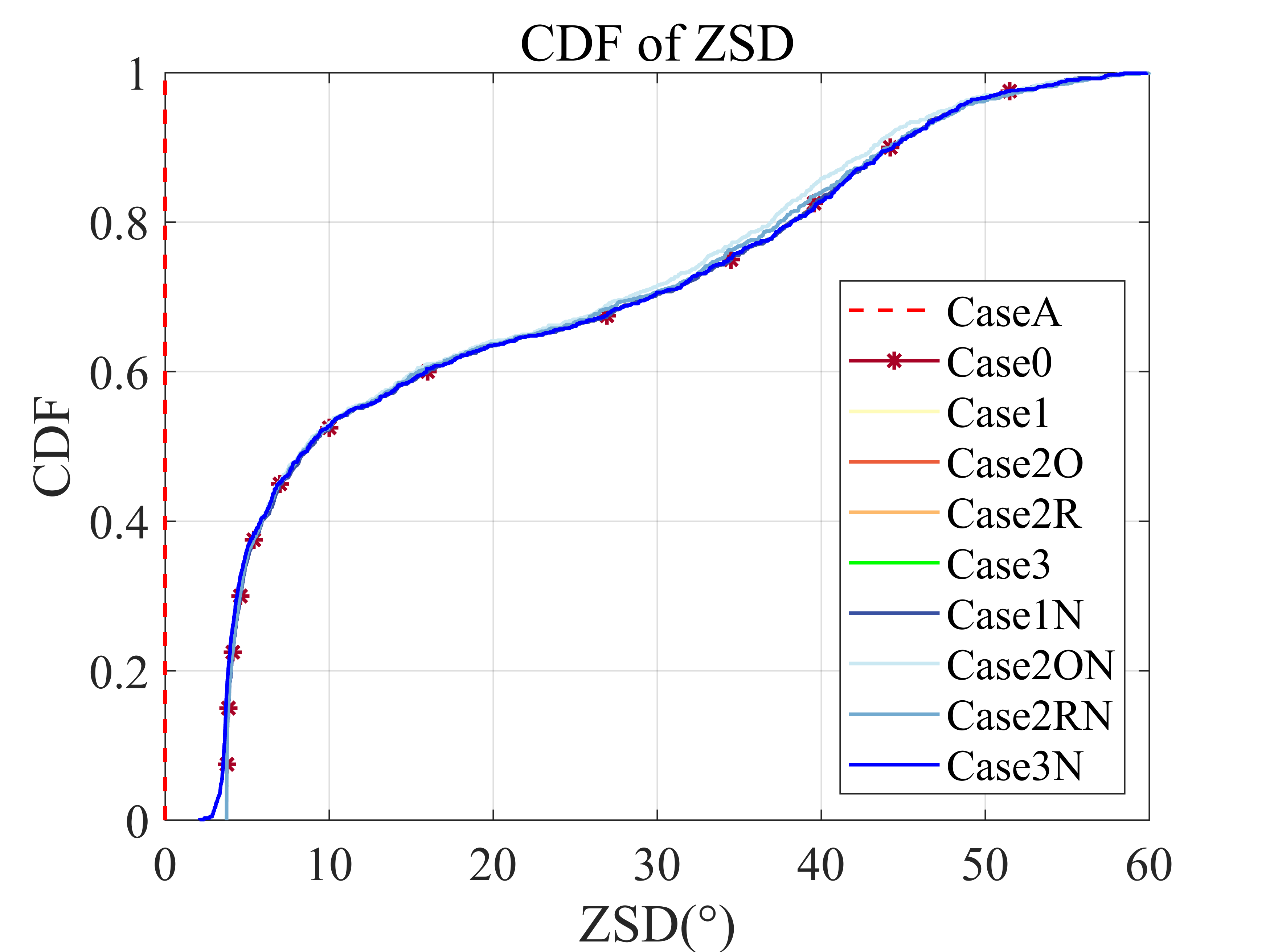}
		\caption*{ZSD of LOS case 3}
	
	\end{minipage}
    \begin{minipage}{0.245\linewidth}
		\centering
		\includegraphics[width=1\linewidth]{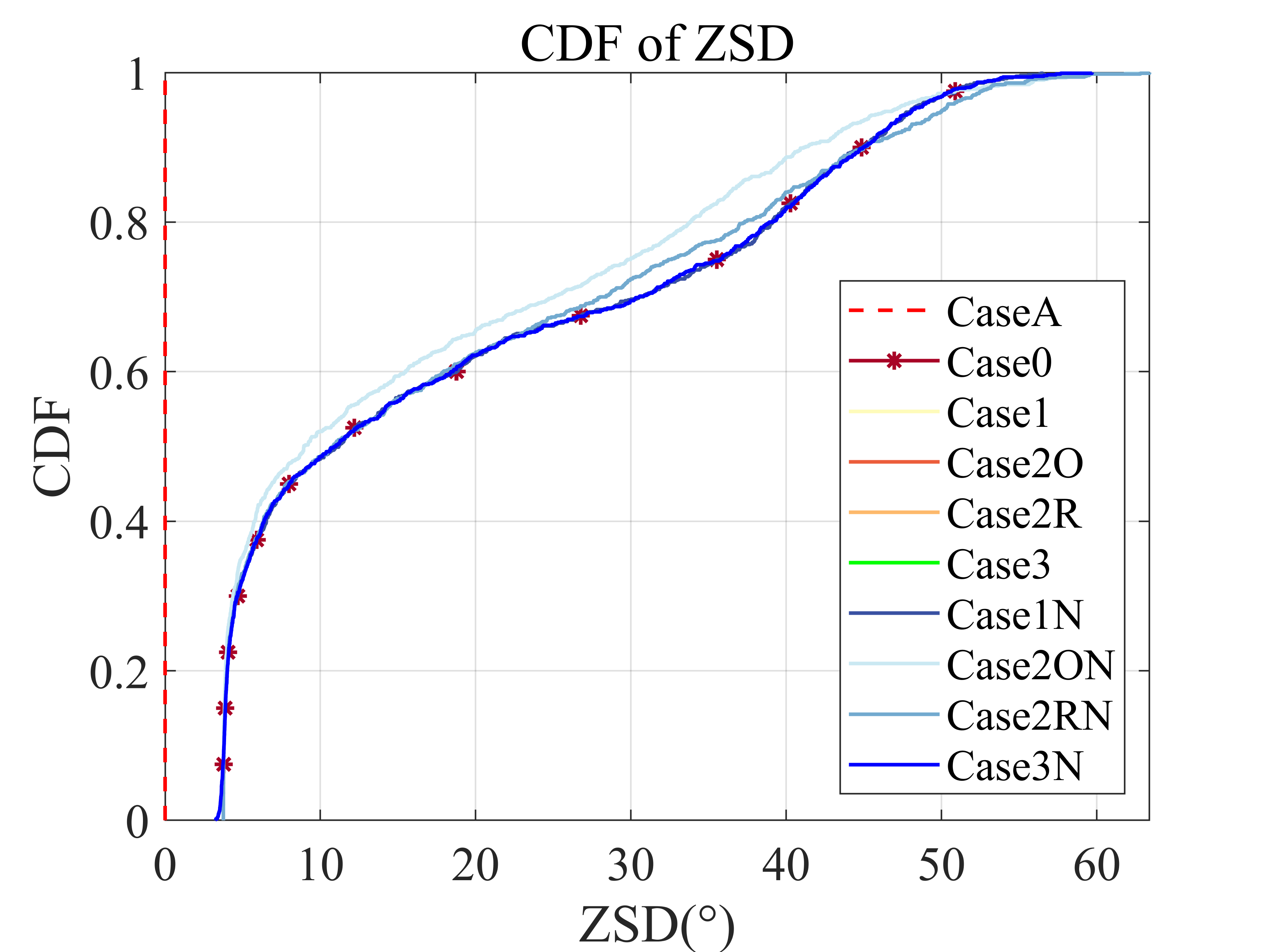}
		\caption*{ZSD of LOS case 4}

	\end{minipage}
	
 \caption{Simulation results for concatenation in UMi scenario }
\label{fig_AS}
\end{figure}

In the simulation, we assume that the total coupling number is determined by selecting the minimum number of clusters when the Tx-target and target-Rx links have differing numbers of clusters (e.g., 12 clusters for UMi-LOS and 19 clusters for UMi-NLOS, selecting the smaller value of 12). As observed in Figure \ref{fig_AS}, Case 1, Case 2O, and Case 2R exhibit a lower power percentage compared to Case 0 (Full Convolution), and overall normalization does not affect the statistical distribution of the spread values (DS and AS). The performance of random coupling at the cluster level (Case 2R) is closer to that of Case 0 than the performance of ordered coupling (Case 2O). 
As observed in LOS case2 and LOS case3 (LOS+NLOS and NLOS+LOS conditions), reserving only the paths associated with LOS paths in Case A results in a zero value for angle spread in the link where the LOS path resides either the Tx-target link or the target-Rx link. Importantly, we observe that the results approximate Case 0 as Case 1N, 2ON, and 2RN by normalizing the power of NLOS-NLOS in Case 1, 2O, and 2R. Furthermore, Case 1N (full convolution at the cluster level followed by normalization) and Case 0 (full convolution) are identical. The CDF curves of case 1N, 2ON, 2RN, and 3N are similar in LOS case 1 ( LOS-LOS condition). Case1N exhibits statistical characteristics identical to Case0 (full convolution), while Case2RN closely approximates the distribution of full convolution than Case2ON.  In conclusion, As for the concatenation of the target channel, normalization of the NLOS-NLOS paths is necessary to maintain the parameter distribution of the channel across different simulation options. Meanwhile, the performance of random coupling in simulation option 2 (Cluster 1-by-1 randomly coupling) is superior to sequential coupling, so random coupling can be prioritized. 

\subsection{RCS measurement and simulation}
\begin{figure}[!h]
\centering
\includegraphics[width=6in]{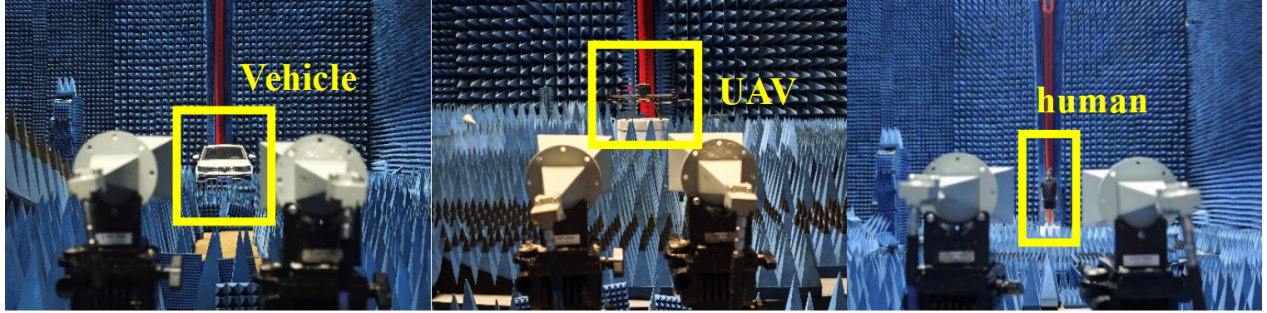}
\caption{RCS measurement scenarios of UAV, Human and Vehicle in an anechoic chamber \cite{28} Copyright [2024] [ARTT-Lab].}
\label{fig_MS}
\end{figure}

We have planned a series of RCS measurement campaigns to acquire data for supporting the precise modeling of target RCS. Vehicles, humans, and UAVs have been selected as the sensing targets for measurement, and these activities will be conducted in an anechoic chamber. An anechoic chamber is utilized to effectively minimize environmental interference. Within this controlled environment, RCS measurements are conducted for various sensing targets, including UAVs, humans, and vehicles. The measurement system comprises two horn antennas, which function as the sensing transmitter and receiver, a Vector Network Analyzer (VNA), a power amplifier, a PC controller, and other components, as illustrated in Figure \ref{fig_MS}. The two horn antennas are positioned in close proximity (with a spacing of approximately 10 cm) to emulate the Mono-static sensing mode, while the sensing target is placed at the center of an automatic turntable. By controlling the rotation of the turntable, the target's echo is measured at different angles. The turntable rotates 360° clockwise in 5° increments. Detailed measurement parameters are provided in Table \ref{tab5}.

\begin{figure}[!h]
	\centering
	\begin{minipage}{0.32\linewidth}
		\centering
		\includegraphics[width=1\linewidth]{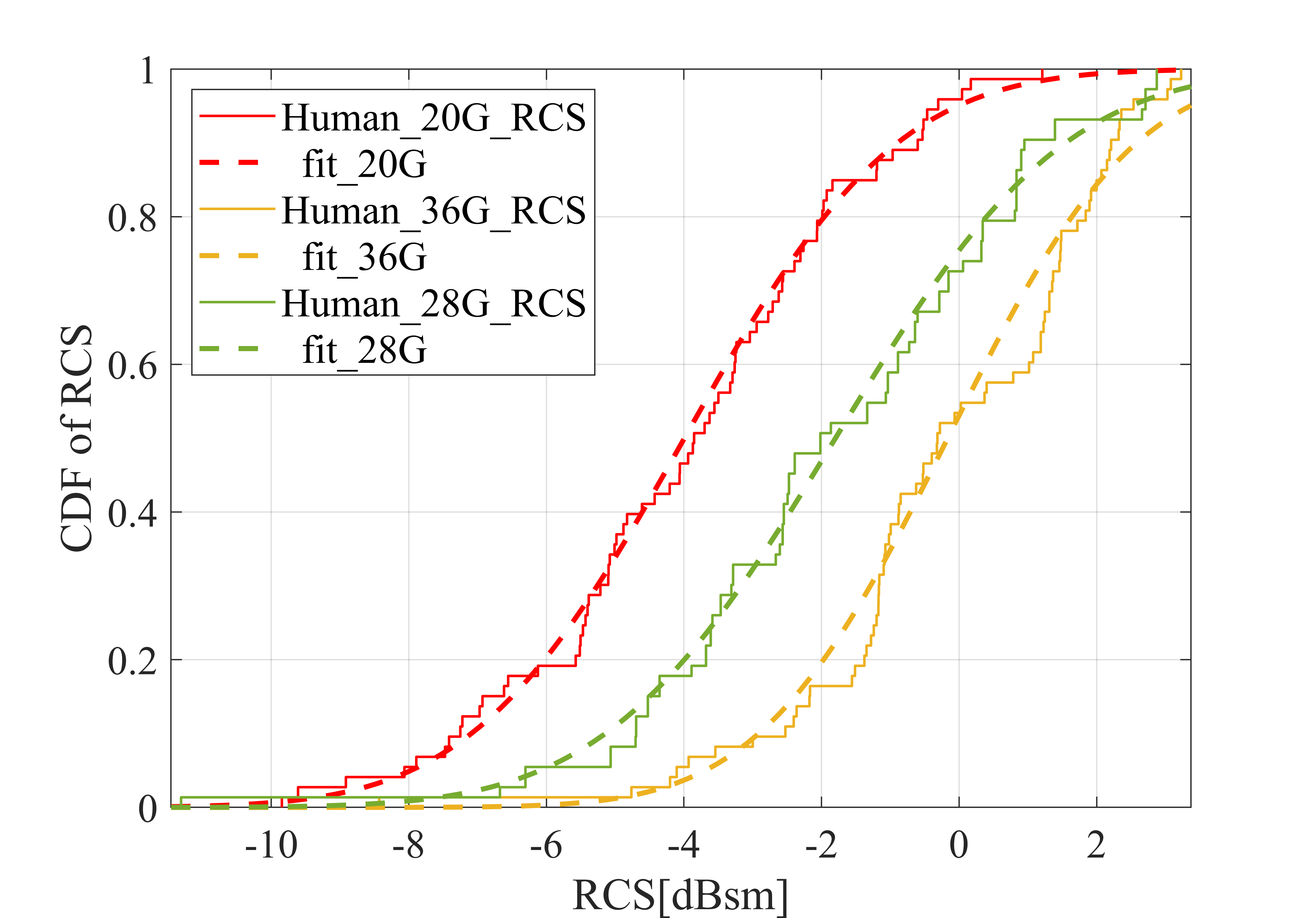}
		\caption*{(a)}
		
	\end{minipage}
	\begin{minipage}{0.32\linewidth}
		\centering
		\includegraphics[width=1\linewidth]{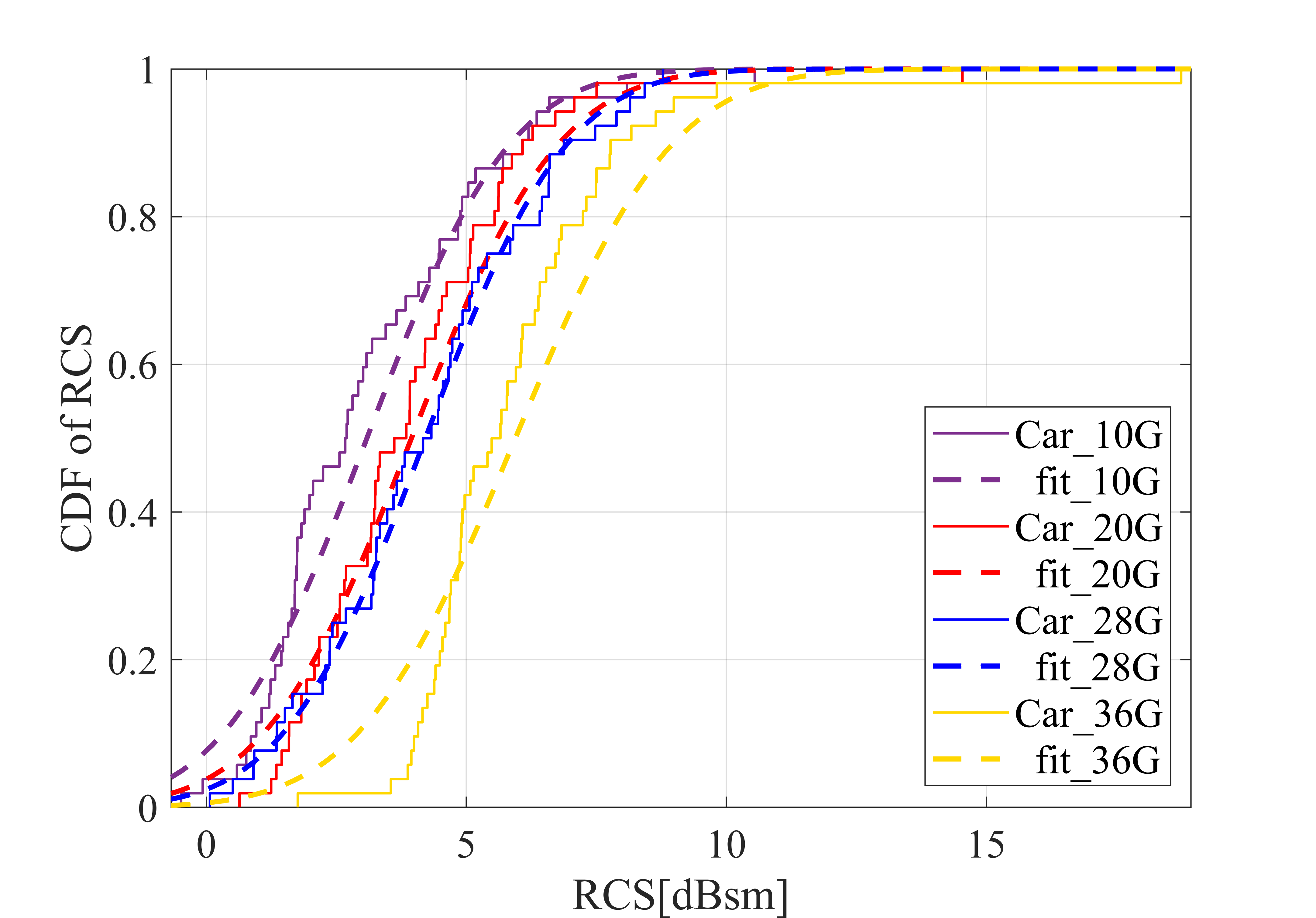}
		\caption*{(b)}
	
	\end{minipage}
    \begin{minipage}{0.32\linewidth}
		\centering
		\includegraphics[width=1\linewidth]{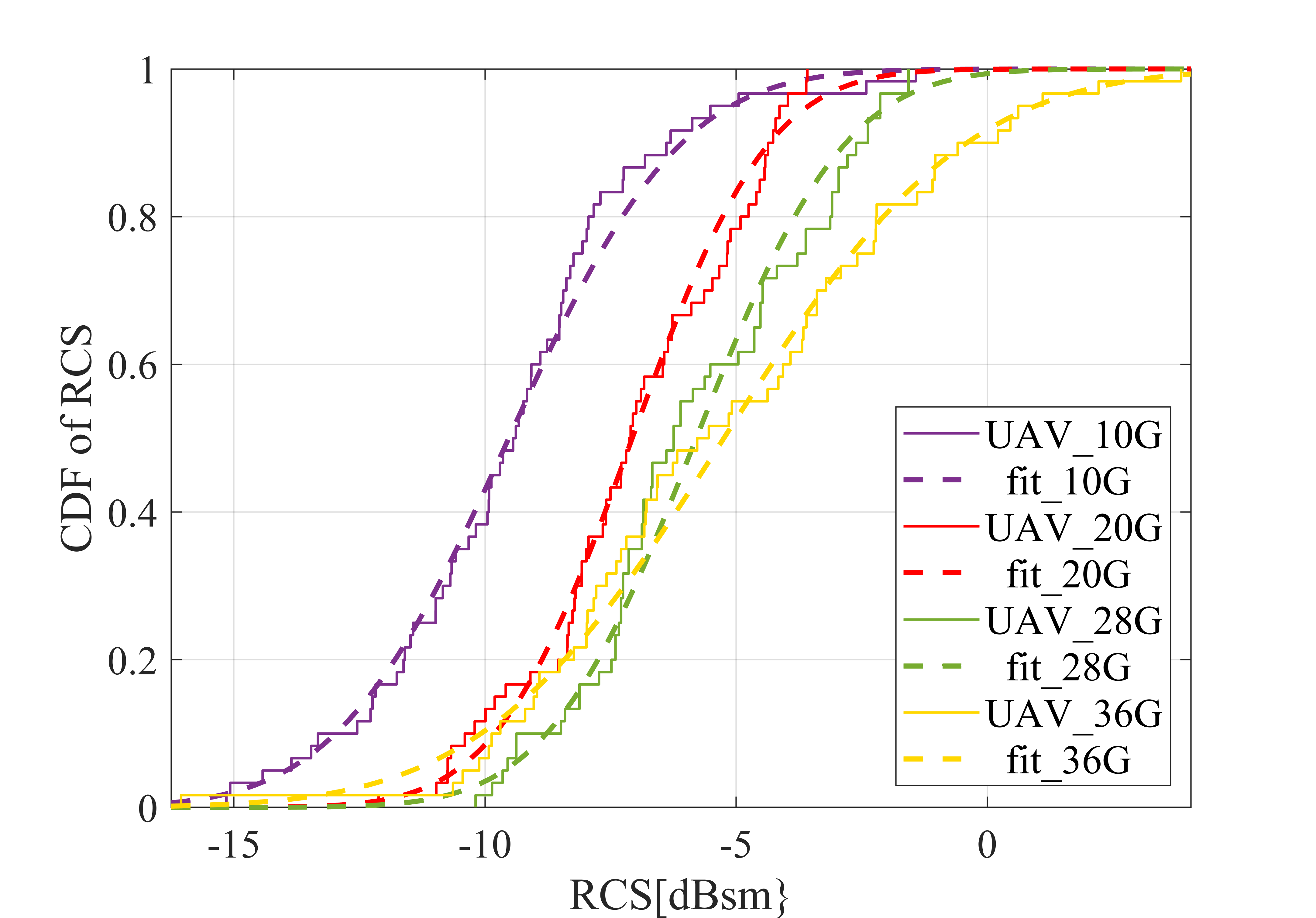}
		\caption*{(c)}
	
	\end{minipage}    
    \caption{CDF of the measured RCS of human, vehicle, and UAV at different frequencies. (a) human. (b) vehicle. (c) UAV.}
 \label{RCS}
\end{figure}

\begin{table}[!h] 
\footnotesize
\caption{Measurement scheme of sensing target RCS}
\label{tab5}
\tabcolsep 60pt 
\begin{tabular*}{\textwidth}{cc}
\toprule
 Parameters & Value/Type \\\hline
 Carrier frequency & 36GHz, 28GHz, 20GHz, 15GHz, 10GHz \\ 
 Bandwidth & 3GHz \\
 Tx/Rx antenna type & Horn/Horn \\
 Polarization & Vertical \\
 Tx/Rx distance form target & 6m, 12m \\
 Sensing mode & Mono-static\\
 Target & Human, Vehicle, UAV \\
\bottomrule
\end{tabular*}
\end{table}
Based on the RCS measurement results, Figure \ref{RCS} presents the cumulative distribution functions (CDFs) of the RCS for humans, vehicles, and UAVs at different frequencies, respectively. Each CDF is fitted using a normal distribution function. During this process, angles exhibiting strong directionality for vehicles and UAVs are excluded, while angles with similar RCS values are retained. Subsequently, a normal distribution function is applied to fit the data, enabling an investigation into the frequency dependence of the sensing targets. In Figure \ref{RCS}, it can be observed that the RCS of the three sensing targets—humans, vehicles, and UAVs—exhibits a certain frequency-dependent characteristic, where the RCS values increase with rising frequency. For the vehicle target, however, there is a noticeable deviation between the fitted normal distribution and the measured values. This discrepancy arises because the vehicle's relatively large size and the significant differences in the surface areas of its four sides introduce a deterministic angle-dependent component to its RCS. Further research is required to explore modeling methods for the deterministic component of the vehicle's RCS. In contrast, the RCS values of human and UAV targets can be well-fitted using a log-normal distribution. Based on the measurement results, we provide reference tables for the fitted $\mu$ and $\sigma$ values of humans and UAVs under various conditions. Due to the extensive data, specific values can be obtained from proposals \cite{27} and \cite{28}.

\section{Conclusion}
In this paper, we extend the 3GPP GBSM standard channel model by incorporating the latest advancements in 3GPP ISAC standardization, thereby enhancing the integrated sensing and communication channel simulation capabilities for 6G. The new model accommodates both Bi-static and Mono-static sensing modes, supporting four types of sensing targets: UAVs, vehicles, humans, and AGVs. This model aligns with prime options outlined in the 3GPP standardization progress, including methods for modeling RCS size and scale, polarization, and a combination of target and background channels.

Through down-selection simulations of the 8 concatenation options provided in the \#118bis meeting, we observed the relation between different cascading methods compared to the fully convoluted baseline. NLOS-NLOS power normalization should be performed, as it ensures overall small-scale power normalization and brings the statistical characteristics of different cascading options closer to Option 0. After NLOS-NLOS power normalization, the statistical distribution of Option 1 matches exactly with that of Option 0. In Option 2, the random matching method performs better than the sequential coupling method. Option 3 demonstrates excellent performance in fitting the fully convoluted result. However, Option 3 abandons the traditional cluster concept. If the cluster concept is necessary (e.g., for MIMO OTA testing), we recommend using Option 2, which retains the cluster concept while closely approximating the fully convoluted results.

This advancement accelerates the standardization of the 6G ISAC channel model and the evaluation of novel algorithms, as it enables the utilization of existing 5G standard parameter tables alongside the proposed cascading model and measured RCS data to simulate ISAC channels. The unified simulation framework applies to all future developments in ISAC channel model standardization, offering insights into the standardization simulation process and facilitating the progress of simulation evaluations.

It is critical to emphasize that the Mono-static background channel cannot be directly modeled by repurposing existing channel modeling methodologies and parameters outlined in current 3GPP technical reports. RAN1 is actively progressing discussions on the development of Mono-static background channel modeling. Future research is required to address the modeling of the Mono-static background channel, as well as the spatial consistency modeling in the ISAC channel. Nevertheless, the proposed method for combining the target channel and background channel in the paper is universally applicable across various scenarios.

%%% Acknowledgements.

\Acknowledgements{This work was supported in part by the National KeyR\&D Program of China (Grant No. 2023YFB2904805), in part by the National Natural Science Foundation of China (Grant Nos. 62341128 and 62201087), in part by the Beijing Natural Science Foundation - Xiaomi Innovation Joint Fund (Grant No. L243002), in part by Guangdong Major Project of Basic and Applied Basic Research (Grant No. 2023B0303000001) and in part by the Beijing University of Posts and Telecommunications-China Mobile Research Institute Joint Innovation Center.}

\end{document}